\documentclass{article}

\usepackage{arxiv}
\usepackage{doi}

\usepackage{fixltx2e}
\newcommand\sub\textsubscript

\usepackage{placeins}

\usepackage[labelfont=bf]{caption}
\usepackage{subcaption}

\usepackage{graphicx}
\graphicspath{{figures/}}

\usepackage{enumitem}
\setlist[itemize]{noitemsep, topsep=0pt}
\setlist[enumerate]{noitemsep, topsep=0pt}

\usepackage{amsmath}
\usepackage{amssymb}
\usepackage{mathtools}
\usepackage{txfonts}
\usepackage{bm}

\DeclareMathOperator{\TP}{TP}
\DeclareMathOperator{\TN}{TN}
\DeclareMathOperator{\FP}{FP}
\DeclareMathOperator{\FN}{FN}
\DeclareMathOperator{\UA}{UA}
\DeclareMathOperator{\PA}{PA}
\DeclareMathOperator{\OA}{OA}
\DeclareMathOperator{\BA}{BA}

\usepackage[table,dvipsnames]{xcolor}

\usepackage{pgf}

\colorlet{worst}{red!30}
\colorlet{mid}{yellow!20}
\colorlet{best}{green!60}

\newcommand*{\MinScore}{0.0}%
\newcommand*{\MidScore}{50.0}%
\newcommand*{\MaxScore}{100.0}%
\newcommand{\scorecolor}[1]{%
    \ifdim #1 pt > \MidScore pt
        \pgfmathsetmacro{\PercentColor}{max(min(100.0*(#1 - \MidScore)/(\MaxScore-\MidScore),100.0),0.00)} %
        \edef\x{\noexpand\cellcolor{best!\PercentColor!mid}}\x{#1\%}
    \else
        \pgfmathsetmacro{\PercentColor}{max(min(100.0*(\MidScore - #1)/(\MidScore-\MinScore),100.0),0.00)} %
        \edef\x{\noexpand\cellcolor{worst!\PercentColor!mid}}\x{#1\%}
    \fi
}

\colorlet{low}{blue!40}
\colorlet{neutral}{white!100}
\colorlet{high}{red!40}

\newcommand*{\MinLag}{-0.5}%
\newcommand*{\MidLag}{0}%
\newcommand*{\MaxLag}{0.5}%
\newcommand{\lagcolor}[1]{%
    \ifdim #1 pt > \MidLag pt
        \pgfmathsetmacro{\PercentColor}{100*max(min((#1 - \MidLag)/(\MaxLag-\MidLag),1.0),0.00)} %
        \edef\x{\noexpand\cellcolor{high!\PercentColor!neutral}}\x{#1}
    \else
        \pgfmathsetmacro{\PercentColor}{100.0*max(min((\MidLag - #1)/(\MidLag-\MinLag),1.0),0.00)} %
        \edef\x{\noexpand\cellcolor{low!\PercentColor!neutral}}\x{#1}
    \fi
}

\usepackage[ruled,vlined,linesnumbered]{algorithm2e}

\usepackage{hyperref}
\hypersetup{
}
\usepackage[round]{natbib}

\usepackage[nameinlink]{cleveref}

\newcommand*{\numnameref}[1]{\hyperref[{#1}]{\ref*{#1} \nameref*{#1}}}
\newcommand*{\fullref}[1]{\hyperref[{#1}]{\cref*{#1} \nameref*{#1}}}
\newcommand*{\Fullref}[1]{\hyperref[{#1}]{\Cref*{#1} \nameref*{#1}}}

\usepackage{booktabs}
\usepackage{tabularx}
\usepackage{makecell}
\usepackage{multirow}
\usepackage{bigdelim}
\usepackage{array}
\usepackage{makecell}
\newcolumntype{P}[1]{>{\raggedright\arraybackslash}p{#1}}
\newcolumntype{C}[1]{>{\centering\arraybackslash}p{#1}}

\usepackage[toc, acronym, nogroupskip]{glossaries}

\glsdisablehyper
\newacronym{AGB}{AGB}{Above Ground Biomass}
\newacronym{UA}{UA}{User's Accuracy}
\newacronym{PA}{PA}{Producer's Accuracy}
\newacronym{OA}{OA}{Overall Accuracy}
\newacronym{BA}{BA}{Balanced Accuracy}
\newacronym{SAR}{SAR}{Synthetic Aperture Radar}
\newacronym{NDVI}{NDVI}{Normalized Difference Vegetation Index}
\newacronym{EVI}{EVI}{Enhanced Vegetation Index}
\newacronym{RMSE}{RMSE}{root mean square error}
\newacronym{MRV}{MRV}{Measurement, Reporting and Verification}
\newacronym{NFMS}{NFMS}{National Forest Monitoring System}
\newacronym{IPCC}{IPCC}{International Panel on Cimate Change}

\makeglossaries

\title{Detecting Deforestation from Sentinel-1 Data in the Absence of Reliable Reference Data}

\author{ \href{https://orcid.org/0000-0003-0743-1332}{\includegraphics[scale=0.06]{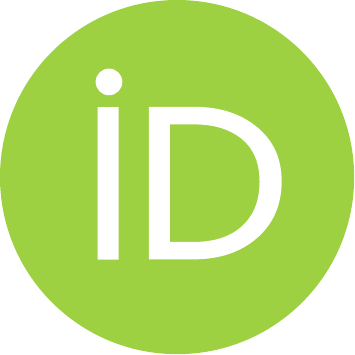}\hspace{1mm}Johannes N.~Hansen}\\
	School of Mathematics\\
	University of Edinburgh\\
	Edinburgh, UK EH9 3FD\\
	\texttt{johannes.hansen@ed.ac.uk} \\
	\And
	\href{https://orcid.org/0000-0002-5690-4055}{\includegraphics[scale=0.06]{orcid.pdf}\hspace{1mm}Edward T.A.~Mitchard} \\
	School of GeoSciences\\
	University of Edinburgh\\
	Edinburgh, UK EH8 3FF\\
	\texttt{edward.mitchard@ed.ac.uk} \\
	\And
	\href{https://orcid.org/0000-0002-4803-1602}{\includegraphics[scale=0.06]{orcid.pdf}\hspace{1mm}Stuart King} \\
	School of Mathematics\\
	University of Edinburgh\\
	Edinburgh, UK EH9 3FD\\
	\texttt{S.King@ed.ac.uk} \\
}

\begin{document}
\maketitle

\begin{abstract}
    Forests are vital for the wellbeing of our planet.
    Large and small scale deforestation
    across the globe is threatening the stability of our climate,
    forest biodiversity,
    and therefore the preservation of fragile ecosystems and our natural habitat as a whole.
    With increasing public interest in climate change issues and forest preservation,
    a large demand for carbon offsetting,
    carbon footprint ratings,
    and environmental impact assessments is emerging.
    Most often, deforestation maps are created from optical data such as Landsat and MODIS.
    These maps are not typically available at less than annual intervals
    due to persistent cloud cover in many parts of the world,
    especially the tropics where most of the world's forest biomass is concentrated.
    Synthetic Aperture Radar (SAR) can fill this gap as it penetrates clouds.
    We propose and evaluate a novel method for deforestation detection
    in the absence of reliable reference data which often constitutes the largest practical hurdle.
    This method achieves a change detection sensitivity
    (producer's accuracy) of 96.5\% in the study area,
    although false positives lead to a lower user's accuracy of about 75.7\%,
    with a total balanced accuracy of 90.4\%.
    The change detection accuracy is maintained when adding up to 20\%
    noise to the reference labels.
    While further work is required to reduce the false positive rate,
    improve detection delay,
    and validate this method in additional circumstances,
    the results show that Sentinel-1 data have the potential
    to advance the timeliness of global deforestation monitoring.
\end{abstract}














\section{Introduction}


Accurate and timely deforestation monitoring is crucial
for enabling effective prevention of deforestation
and the protection of existing forest reserves.
The United Nations Framework Convention on Climate Change (UNFCC)
has adopted the REDD+ (Reducing emissions from deforestation and forest degradation)
rulebook to emphasize the importance of forest conservation
for the halting and reversal of climate change.
Developing countries under this framework are required to implement \gls{MRV} systems
as part of a \gls{NFMS}.
The \gls{IPCC} dictates quality standards
for these monitoring systems \citep{IPCC2003}.
In particular, they must be accurate, scalable, trusted, and comparable.
Furthermore, these systems are required to use remote sensing,
but there are no specifications regarding the exact technologies and methods.
\citet{Reiche2016} note that the majority of REDD+ member countries
in the tropics are using Landsat, but no \gls{SAR} data,
for their respective \glspl{MRV}.
As most of these countries are subject to 70--80\% cloud cover throughout the year,
the use of optical data results in significant detection delays and potential blind spots.
This problem can be alleviated by using \gls{SAR} data
which is unaffected by clouds at C-band or longer \citep{Lu2006}.

The capabilities of \gls{SAR} for deforestation monitoring
are widely demonstrated
\citep{Mitchard2011,Bouvet2018,Rahman2010,Soja2018,Hansen2020}
and \citet{Lehmann2015} have presented a strategy for
combining \gls{SAR} and optical data in a large-scale
forest monitoring system.
The lack of \gls{SAR} based national monitoring systems
may be due to the relative ease of optical data processing,
the larger historic archives,
and the comparatively limited availability of free \gls{SAR} data and \gls{SAR} processing tools
\citep{Reiche2016}.

One of the biggest challenges in deforestation monitoring
is the (un)availability of good reference data.
Many detection algorithms rely on some form of
supervised machine learning,
and any such algorithm is only as good as its training data.
The difficulty in acquiring these reference data
lies in the global scale of the phenomenon,
its temporal sensitivity
(many detection methods may fail even a few weeks
after deforestation as some vegetation has regrown),
local political tension including armed conflicts,
the remote location of some affected areas,
and many more \citep{RodriguezVeiga2017}.

In practice,
most reference data are generated by manually
drawing polygons around areas that look like deforestation
in high-resolution aerial or satellite imagery.
Generally, there is a trade-off between spatial and temporal cover
for any reference data set.
Temporally dense data are only available over very few selected monitoring sites
that may or may not be representative of forest elsewhere.
Reference data of forest degradation (such as selective logging)
are even more difficult to obtain.
This is problematic, as according to some sources,
forest degradation now accounts for more loss of biomass than deforestation
\citep{Qin2021}.
The most widely used global data set
is published at \emph{GlobalForestWatch} (\url{http://globalforestwatch.org/}).
In a highly cited paper,
\citet{Hansen2013} present findings from a global survey of
annual forest cover, gain, and loss
for the years 2000 to 2012.
Their analysis is based on Landsat data,
resulting in a global spatial resolution of 30\,m.
However, these data are not suitable for use as training data
for machine learning algorithms,
as they are themselves the output of a classification algorithm.
The lack and quality of reference data
remains an issue to be solved for deforestation monitoring.
Ideally,
the exact time and type of change is known for a large area.
However,
data of this kind are as of yet infeasible
because they require continuous large scale monitoring.
For global coverage,
the best available forest maps are still only updated at annual intervals,
including
GlobalForestWatch \citep{Hansen2013},
JRC Tropical Moist Forests \citep{Vancutsem2021},
and a global forest map based on ALOS PALSAR data \citep{Shimada2014}.

In order to circumvent the reliance on reference data
to train deforestation detection algorithms,
one could simply apply change detection algorithms
that do not rely on training labels.
However,
this approach has the downside that the detected changes
do not necessarily correspond to deforestation
but may instead reflect land cover transitions
other than forest to non-forest,
or indeed measurement changes that do not represent
a change in the underlying land cover at all,
for example: (1) seasonal changes in vegetation,
(2) growth and harvest cycles in agriculture, or
(3) soil moisture changes due to rainfall.
This is because change detection algorithms are in no way
specific to any particular type of change but merely
pick up statistically significant changes in the raw data,
whether or not these changes correspond to a change in the underlying state.

\citet{Mitchell2017} present an overview of existing
forest degradation monitoring techniques.
The authors of the review
discriminate between two main approaches:
(a) assessment of degradation
via change in canopy cover or proxies (e.g. roads, log decks),
and (b) direct quantification of loss in \gls{AGB}.
The mapping of proxies,
such as the progression of forest roads or
forest fragmentation,
could serve as a risk assessment
for potential degradation
when combined with a proximity metric.
In order to monitor changes in any of these metrics,
a number of change detection algorithms
have been developed specifically
for application in remote sensing data, including
BFAST \citep{Verbesselt2010a},
LandTrendr \citep{Kennedy2010},
CMFDA \citep{Zhu2012b},
and CCDC \citep{Zhu2014a}.
Vegetation indices such as \gls{NDVI} and \gls{EVI},
as well as \gls{SAR} backscatter
are commonly used to monitor forest canopy change.

\citet{Hamunyela2017} note that the trade-off
between spatial and temporal accuracy
in deforestation detection is limiting
the overall accuracy of disturbance maps.
If short temporal detection delays are to be achieved,
this typically results in low omission error
but very high commission error.
In particular,
the use of change magnitude thresholds
is seen as problematic
as it relies on the assumption
that true and false detections
do indeed exhibit different change magnitudes.
While this assumption may hold
for large scale clear-felling,
small scale or scattered forest disturbances may go undetected.

From these considerations we define the basic research question
to be addressed in this paper.
The goal is to develop a robust deforestation detection method
from multivariate and multi-source data.
The method should have a number of qualities:
(1) It should be physics-agnostic to a reasonable degree,
such that prior knowledge about the input data may improve the result,
but is not required.
(2) It should provide reliable deforestation alerts even
in the presence of sub-optimal reference data.
(3) It should scale well with increasing dataset size.
The potential for Sentinel-1 data to distinguish between
forest and non-forest in a wide range of biomes
has already been demonstrated in \citet{Hansen2020}.
In this paper,
we are expanding on that study by proposing a method for detecting changes at a sub-annual timeliness.

\FloatBarrier
\section{Methods}


The deforestation detection problem can be reframed
in several different ways.
For example,
it can be thought of as a classification problem,
where each pixel is classified as forest or non-forest at each time step.
Deforestation events would then be declared
wherever a pixel changes class from forest to non-forest
between consecutive time steps.
This approach has several downsides:
firstly, it requires reliable reference data
that are both temporally and spatially accurate
for training the classifier.
In addition,
it is not very robust because different land cover classes
may look very similar to forest at some time steps,
either because of random fluctuations
(speckle, noise, etc.)
or because of seasonal effects
(greening and senescence in forests;
growth and harvest in agriculture).
This approach would therefore likely detect a multitude of changes
back and forth between classes.

Alternatively,
we can directly apply off-the-shelf change detection algorithms to
the time series data.
This is problematic for similar reasons as the previous approach,
mainly because a change in the time series behavior
does not necessarily indicate a change in land cover,
and vice versa.

In our case,
we want to think about a deforestation event
as a deviation of the pixel from the reference class (stable forest).
We therefore want to detect the point in time
when the pixel time series starts to be
dissimilar to the characteristic forest time series.
This requires a time-dependent measure of similarity between the two time series.
Because of this,
we convert the original time series into
another time series that characterizes how
dissimilar the pixel is to forest at that time.
The original change detection problem is then transformed
into the task of detecting an increase in forest dissimilarity.

\subsection{Study Site and Reference Data}

\begin{figure}[!ht]
    \centering
    \begin{subfigure}[t]{0.24 \textwidth}
        \includegraphics[width=\textwidth]{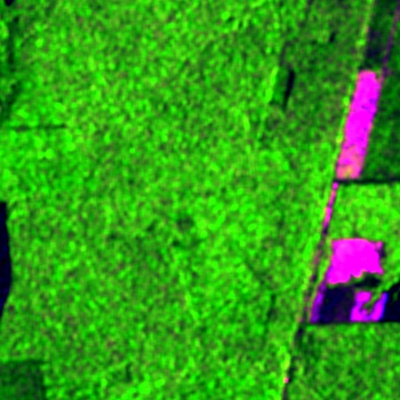}
        \caption{05/01/2017}
    \end{subfigure}
    \begin{subfigure}[t]{0.24 \textwidth}
        \includegraphics[width=\textwidth]{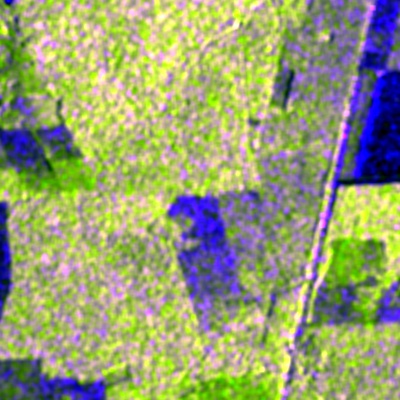}
        \caption{29/05/2017}
    \end{subfigure}
    \begin{subfigure}[t]{0.24 \textwidth}
        \includegraphics[width=\textwidth]{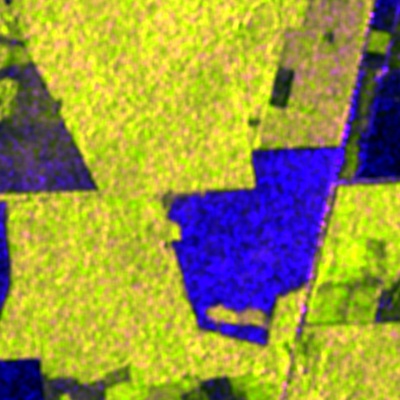}
        \caption{25/11/2017}
    \end{subfigure}
    \begin{subfigure}[t]{0.24 \textwidth}
        \includegraphics[width=\textwidth]{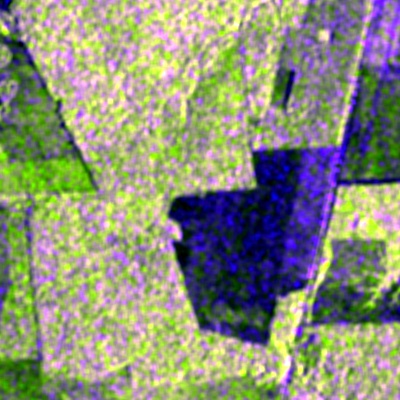}
        \caption{05/06/2018}
    \end{subfigure}
    \begin{subfigure}[t]{0.24 \textwidth}
        \includegraphics[width=\textwidth]{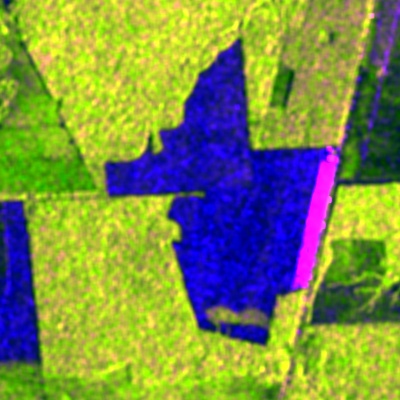}
        \caption{02/12/2018}
    \end{subfigure}
    \begin{subfigure}[t]{0.24 \textwidth}
        \includegraphics[width=\textwidth]{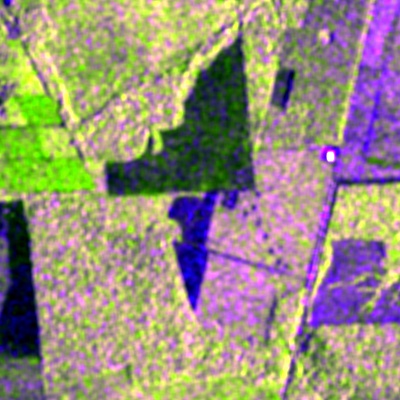}
        \caption{31/05/2019}
    \end{subfigure}
    \begin{subfigure}[t]{0.24 \textwidth}
        \includegraphics[width=\textwidth]{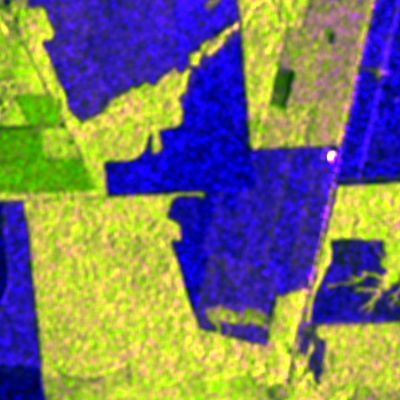}
        \caption{27/11/2019}
    \end{subfigure}
    \begin{subfigure}[t]{0.24 \textwidth}
        \includegraphics[width=\textwidth]{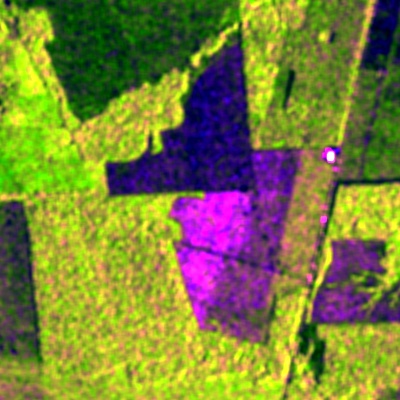}
        \caption{25/05/2020}
    \end{subfigure}
    \caption{This figure shows a series of Sentinel-1 images of the study site in Paragominas, Brazil,
    during the period 2017--2020.
    The RGB channels represent the backscatter in VV, VH, as well as their ratio VV/VH.
    }
    \label{fig:paragominas_s1}
\end{figure}
\begin{figure}[!ht]
    \centering
    \begin{subfigure}[t]{0.24 \textwidth}
        \includegraphics[width=\textwidth]{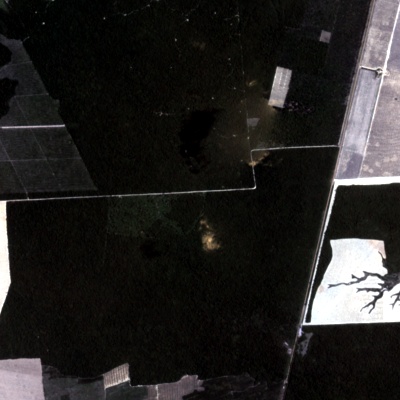}
        \caption{11/2016}
        \label{fig:paragominas_nicfi_2016_11}
    \end{subfigure}
    \begin{subfigure}[t]{0.24 \textwidth}
        \includegraphics[width=\textwidth]{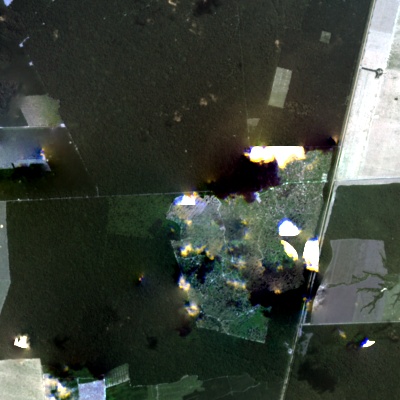}
        \caption{05/2017}
        \label{fig:paragominas_nicfi_2017_05}
    \end{subfigure}
    \begin{subfigure}[t]{0.24 \textwidth}
        \includegraphics[width=\textwidth]{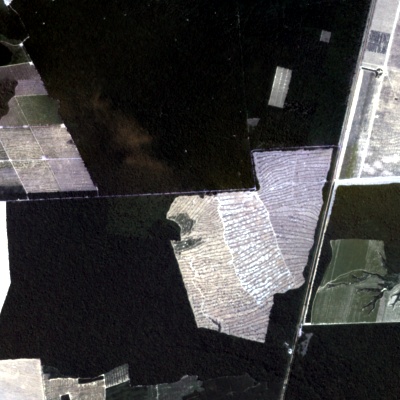}
        \caption{11/2017}
        \label{fig:paragominas_nicfi_2017_11}
    \end{subfigure}
    \begin{subfigure}[t]{0.24 \textwidth}
        \includegraphics[width=\textwidth]{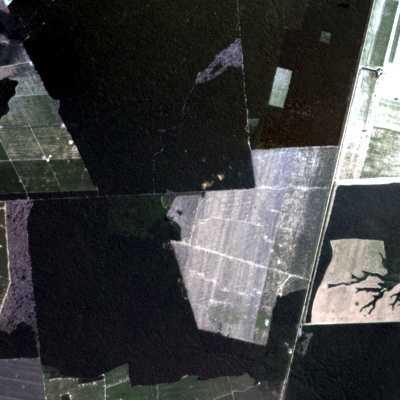}
        \caption{05/2018}
        \label{fig:paragominas_nicfi_2018_05}
    \end{subfigure}
    \begin{subfigure}[t]{0.24 \textwidth}
        \includegraphics[width=\textwidth]{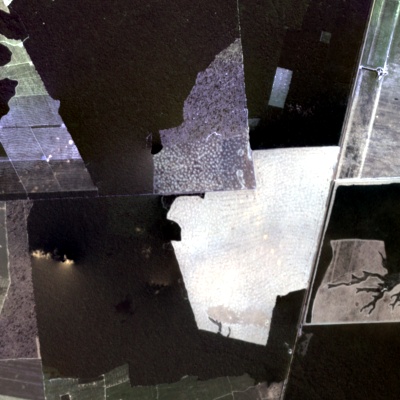}
        \caption{11/2018}
        \label{fig:paragominas_nicfi_2018_11}
    \end{subfigure}
    \begin{subfigure}[t]{0.24 \textwidth}
        \includegraphics[width=\textwidth]{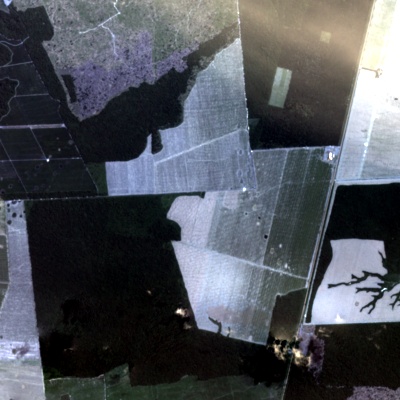}
        \caption{05/2019}
        \label{fig:paragominas_nicfi_2019_05}
    \end{subfigure}
    \begin{subfigure}[t]{0.24 \textwidth}
        \includegraphics[width=\textwidth]{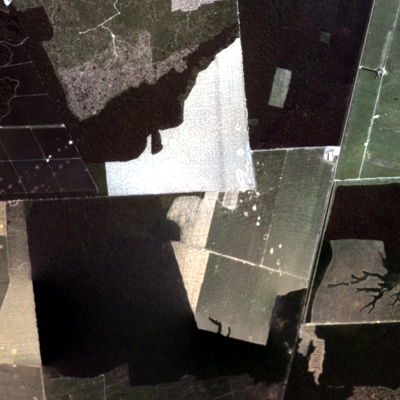}
        \caption{11/2019}
        \label{fig:paragominas_nicfi_2019_11}
    \end{subfigure}
    \begin{subfigure}[t]{0.24 \textwidth}
        \includegraphics[width=\textwidth]{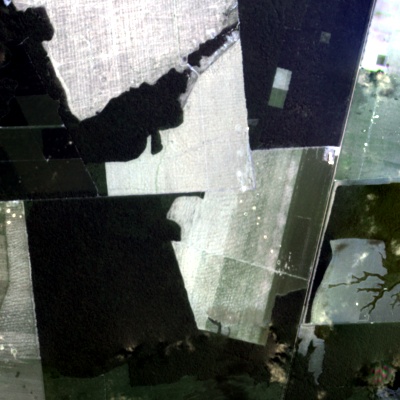}
        \caption{05/2020}
        \label{fig:paragominas_nicfi_2020_05}
    \end{subfigure}
    \caption{This figure shows a series of six-monthly RGB composites
    of the study site in Paragominas, Brazil,
    during the period 2017--2020.
    The images are provided by Planet via the NICFI program \citep{NICFI}.
    }
    \label{fig:paragominas_nicfi}
\end{figure}

For developing and testing the method
we chose a small site near Paragominas in Pará, Brazil,
that exhibits very clear deforestation in large patches
for the years 2017, 2018, and 2019.
\Cref{fig:paragominas_s1} shows a series of Sentinel-1 images
for this study site,
in approximately six-month intervals.
In addition,
\cref{fig:paragominas_nicfi} shows high-resolution optical data for the same area,
provided free of charge by Planet through the NICFI program \citep{NICFI}.
The dates shown for the Sentinel-1 images were chosen to approximately correspond
to the available NICFI images.


\begin{figure}[!ht]
    \centering
    \begin{subfigure}[t]{0.32\textwidth}
        \includegraphics[width=\textwidth]{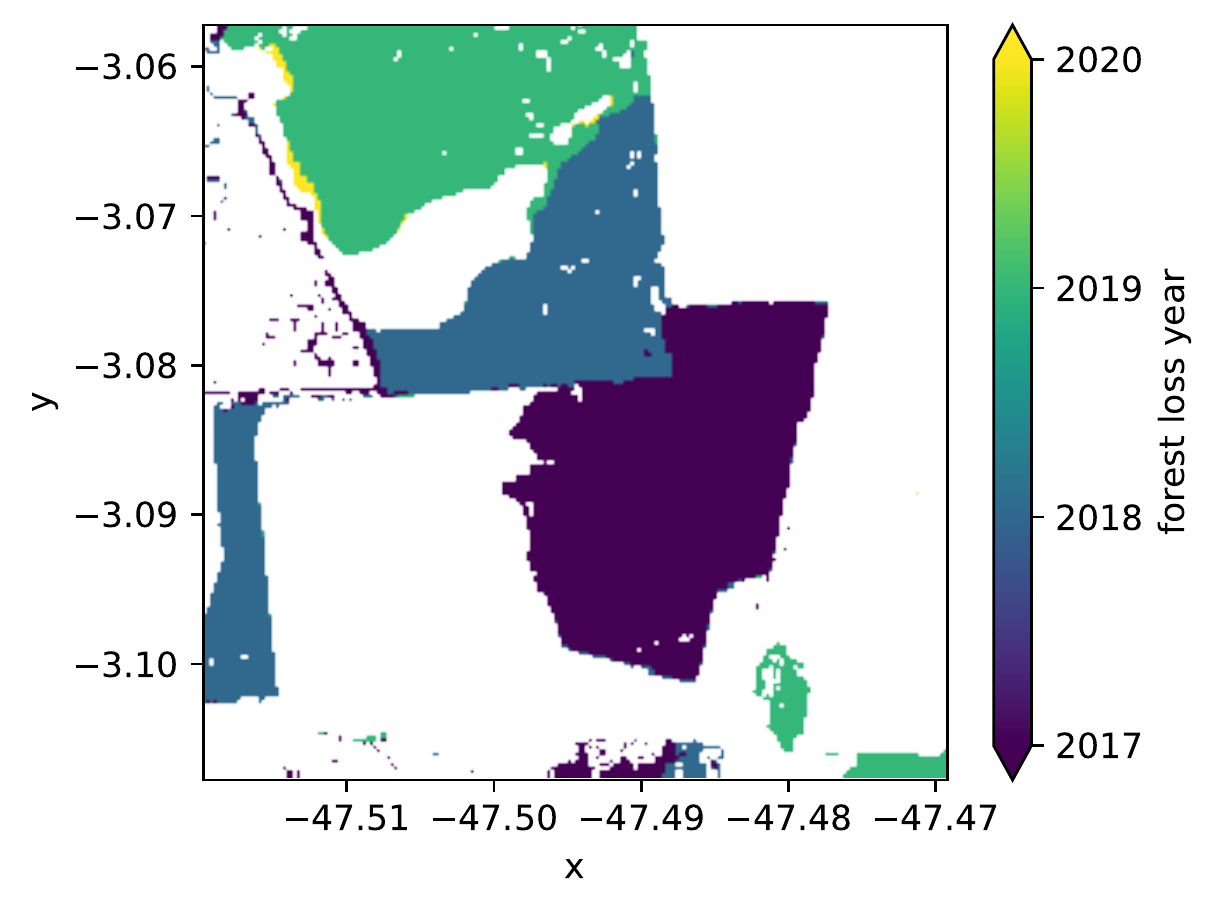}
        \caption{Hansen reference}
        \label{fig:reference_hansen}
    \end{subfigure}
    \begin{subfigure}[t]{0.32\textwidth}
        \includegraphics[width=\textwidth]{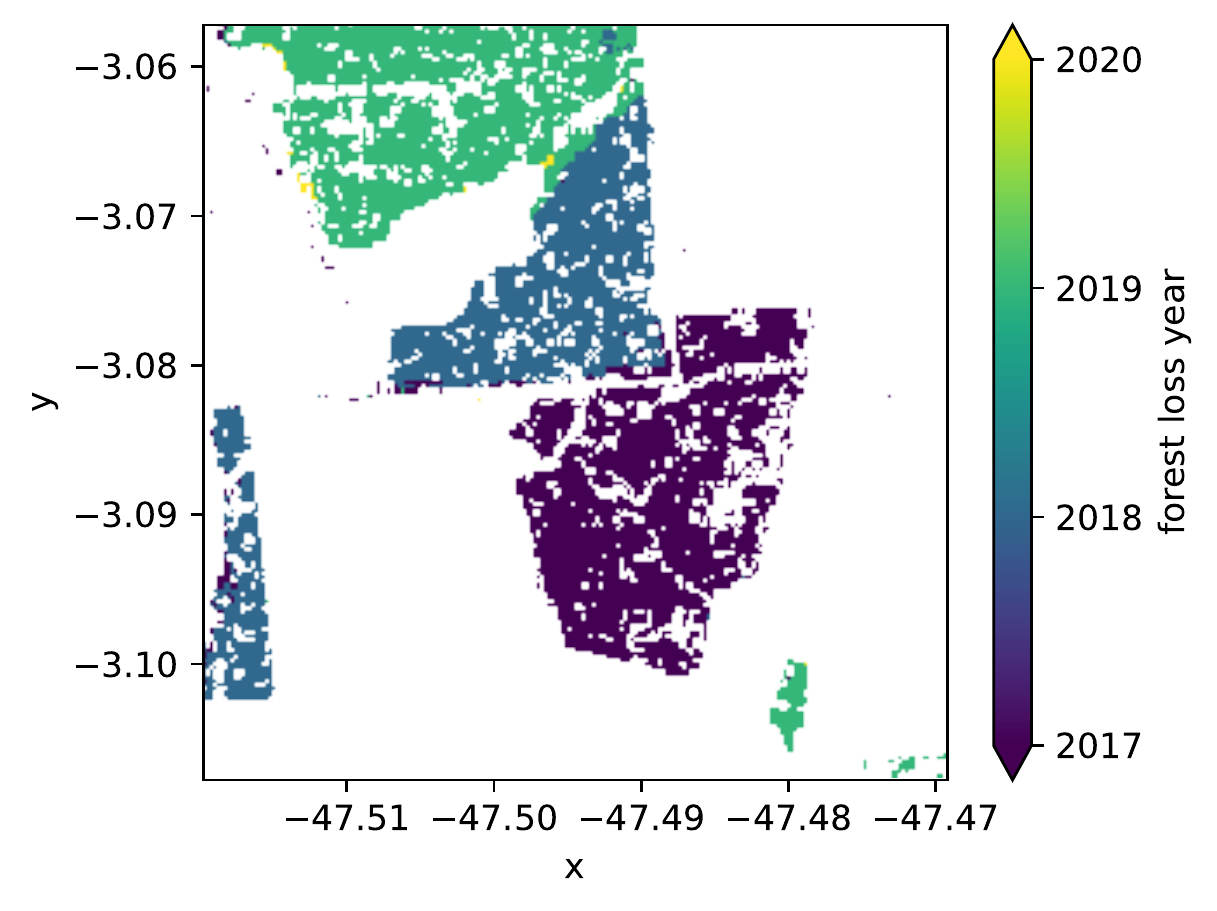}
        \caption{JRC TMF reference}
        \label{fig:reference_jrc}
    \end{subfigure}
    \begin{subfigure}[t]{0.32\textwidth}
        \includegraphics[width=\textwidth]{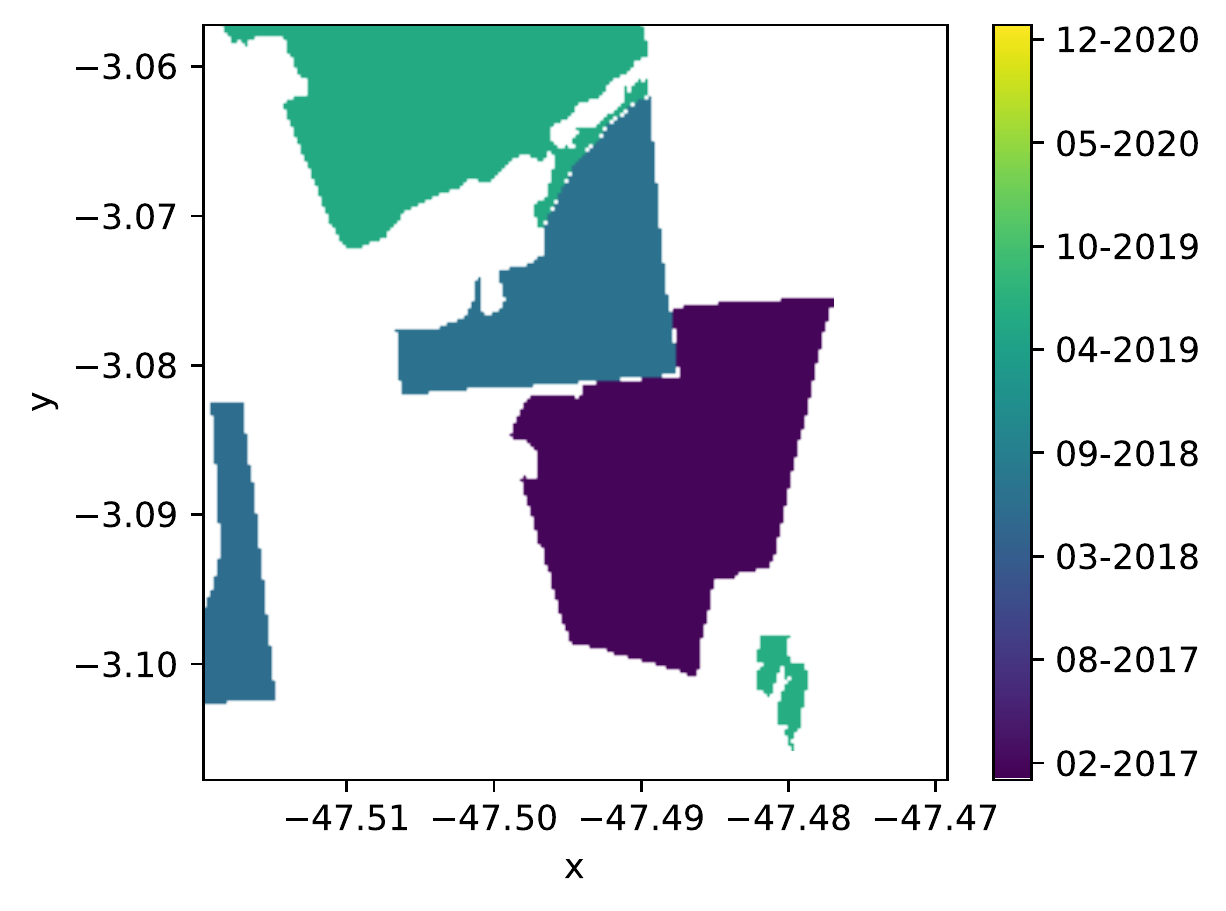}
        \caption{Visually interpreted reference}
        \label{fig:reference_visual}
    \end{subfigure}
    \caption{Reference change maps.
    The colored pixels indicate the year of forest loss,
    whereas white pixels indicate no change,
    which includes stable forest as well as non-forest.}
    \label{fig:reference_change}
\end{figure}

\Cref{fig:reference_change} shows the annual forest loss events
for the study site
obtained from two reference data sets:
(1) The GlobalForestWatch dataset \citep{Hansen2013} and
(2) The Tropical Moist Forests product provided by the EC JRC \citep{Vancutsem2021}.
In addition,
\cref{fig:reference_visual} shows a manual segmentation of the study site
into polygons that exhibit roughly synchronous behavior
throughout the time period.
This way, one date of change can be assigned to each area,
although some of these areas are deforested in a rainbow pattern,
i.e. the change spans a period of time that typically covers a few Sentinel-1 images.
The change dates shown in the figure were derived
from visual interpretation of Landsat-8 data and represent
the latest possible date that the deforestation has likely occurred.
Because of extensive cloud cover in the area,
it is possible that the forest loss happened earlier but
could not be detected until the date shown.
The change detection results will be evaluated with respect to these three maps.

\subsection{Change Detection}

As described earlier,
the accuracy of deforestation detection methods
is hindered by the scarcity and low quality of reference data,
as well as the fact that traditional change detection methods
may pick up changes corresponding to events other than forest loss.
To solve these problems,
and to mitigate the disadvantages of both
the fully supervised and the fully unsupervised methods,
in this paper we are demonstrating a partially supervised approach:
Instead of requiring reference data that capture
land cover changes and therefore need a temporal component,
we have developed a method that only relies on a stationary
forest map and thus classifies pixels as either being
stable forest or not.
Pixels that are not stable forest could
include agriculture,
urban areas,
other forms of vegetation, etc.,
or pixels that undergo deforestation at some point.

This reference forest map can then be used
to detect pixels that deviate from the reference class.
This is done by computing a similarity measure between
the time series of the pixel
and the reference forest time series.
The similarity can then be computed over time
and compared to the expected similarity
if the pixel was forest.
Transitions from forest to non-forest
can then be recorded as a drop in forest similarity
above a set threshold.

The changepoint detection is simplified by the fact
that there is only one type of change of interest
--- forest to non-forest ---
and that there is a maximum of one such change in the time series.
We therefore do not need to worry about detecting multiple changes.
The change detection method is based on the following steps:

\begin{enumerate}
    \item Obtain an initial reference forest mask.
    This could be based on \citet{Hansen2013} or any other forest dataset.
    It does not need to be exhaustive or multitemporal 
    but should cover areas that remain stable forest throughout the time period.

    \item Based on the data and the reference mask,
    compute the reference data distribution for forested areas at each time step by creating a histogram
    of the masked data.
    Let $b(x_{i,t})$ denote the histogram bin
    and $C_{b(x_{i,t})}$ the histogram count at that bin.

    \item From the forest histogram,
    compute the forest similarity for each pixel at each time step $t$, for each variable $i$:

    \begin{equation}
        p_{i,t}(x_{i,t}) = \frac{1}{\sum{C}} C_{b(x_{i,t})}
    \end{equation}

    with $p_{i,t}$ being the forest similarity at time step $t$,
    given the value $x_{i,t}$ for variable $i$.
    In this case,
    the variables are the co-polarized and cross-polarized backscatter (VV and VH).
    However,
    the method generalizes to any multivariate data.

    \item To combine the similarities from all available variables
    into a univariate time series,
    we compute the joint forest similarity as:
    \begin{equation}
        p_{t}(x_t) = \prod_i{p_{i,t}(x_t)}
    \end{equation}

    \item
    For comparison with the expected similarity values if a pixel is forest,
    we compute the same joint similarity for forest pixels and use the $q$-th quantile $p^{F,q}_t$
    as threshold for declaring a pixel non-forest
    (this means that a fraction $q$ of forest pixels would also be declared non-forest at this stage).

    \item Compute the cumulative product of the similarity threshold ratio:
    \begin{align}
        \Lambda_{t}(x) = \begin{cases}
            1 & \text{if} \quad p_{t-1}(x) < p^{F,q}_{t-1} \\
            {p_0(x)}/{p^{F,q}_0} & \text{else if} \quad t = 0 \quad \\
            P_{t-1}(x) \cdot {p_t(x)}/{p^{F,q}_t} & \text{else} \\
        \end{cases}
    \end{align}
    $\Lambda_t(x)$ increases as consecutive time steps fall above the threshold $p^{F,q}_{t}$
    and is therefore a measure of confidence that the pixel is in fact non-forest at this time step.

    \item Set a threshold $L$ for $\Lambda_t$ for the pixel to be considered non-forest.

    \item Finally, a deforestation event is declared at time step $T$
    if $\Lambda_T(x) \ge L$ and $\Lambda_{t}(x) < L \;\forall t < T$.

\end{enumerate}

\begin{figure}[!htb]
    \centering
    \begin{subfigure}[c]{0.3 \textwidth}
        \includegraphics[width=\textwidth]{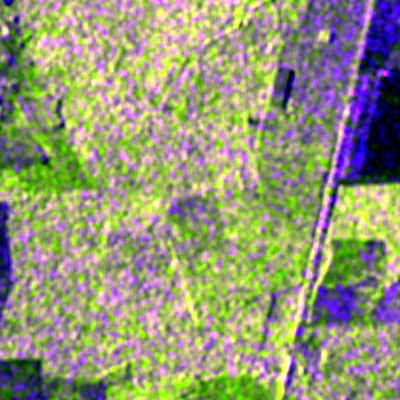}
        \caption{Data}
        \label{fig:demo_image}
    \end{subfigure}
    \begin{subfigure}[c]{0.3 \textwidth}
        \includegraphics[width=\textwidth]{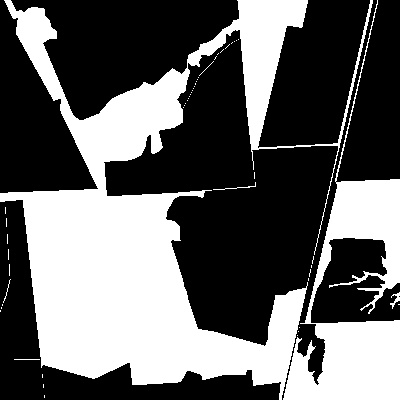}
        \caption{Forest mask}
        \label{fig:demo_mask}
    \end{subfigure}
    \begin{subfigure}[c]{0.35 \textwidth}
        \includegraphics[width=\textwidth]{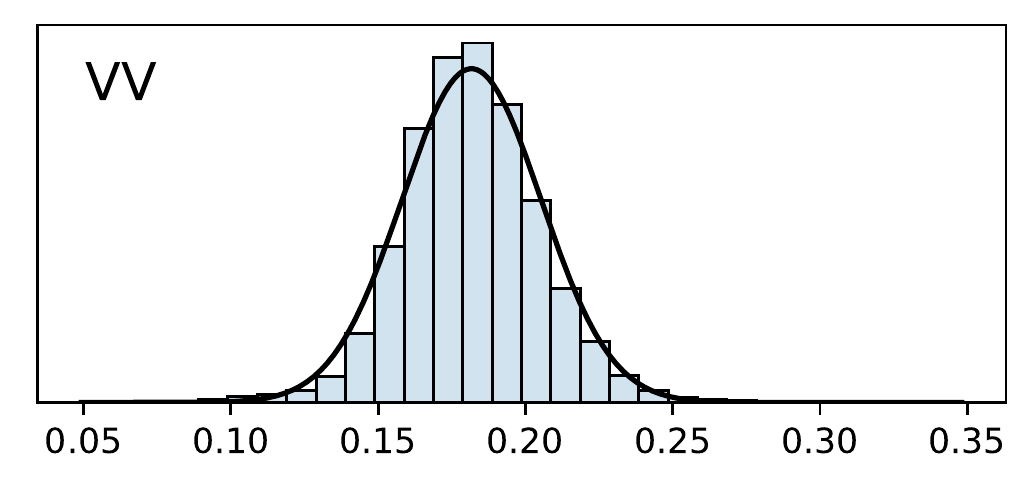}
        \includegraphics[width=\textwidth]{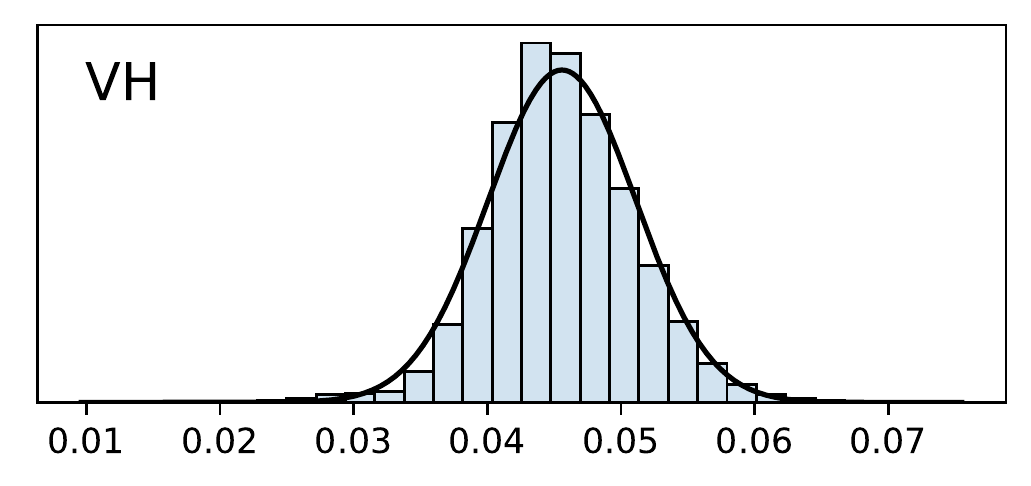}
        \caption{Forest distribution}
        \label{fig:demo_hist}
    \end{subfigure}

    \caption{This figure shows a time slice of Sentinel-1 data for the study site in Paragominas, Brazil
    (\cref{fig:demo_image}),
    a mask showing the areas of stable forest throughout the study period (\cref{fig:demo_mask}),
    and the distribution of VV and VH values for the forest class based on the forest mask (\cref{fig:demo_hist}).}
    \label{fig:demo_distribution}
\end{figure}

For noise reduction,
a non-local means filter was applied to the spatio-temporal datacube \citep{Buades2011}
(implemented in \citet{Hansen2022}).
\Cref{fig:demo_distribution} demonstrates how an initial forest mask can be used to compute
the distribution of feature values for the forest class (steps 1 and 2).
The forest mask for this demonstration was obtained from visual interpretation (see \cref{fig:reference_polygons}),
but would in general be an existing forest map of uncertain quality.
In this case,
the distribution of forest backscatter values looks nearly Gaussian
(as indicated by the Gaussian curve fit),
but this need not be the case in general.

\begin{figure}[!h]
    \centering
    \begin{subfigure}[t]{0.45 \textwidth}
        \includegraphics[width=\textwidth]{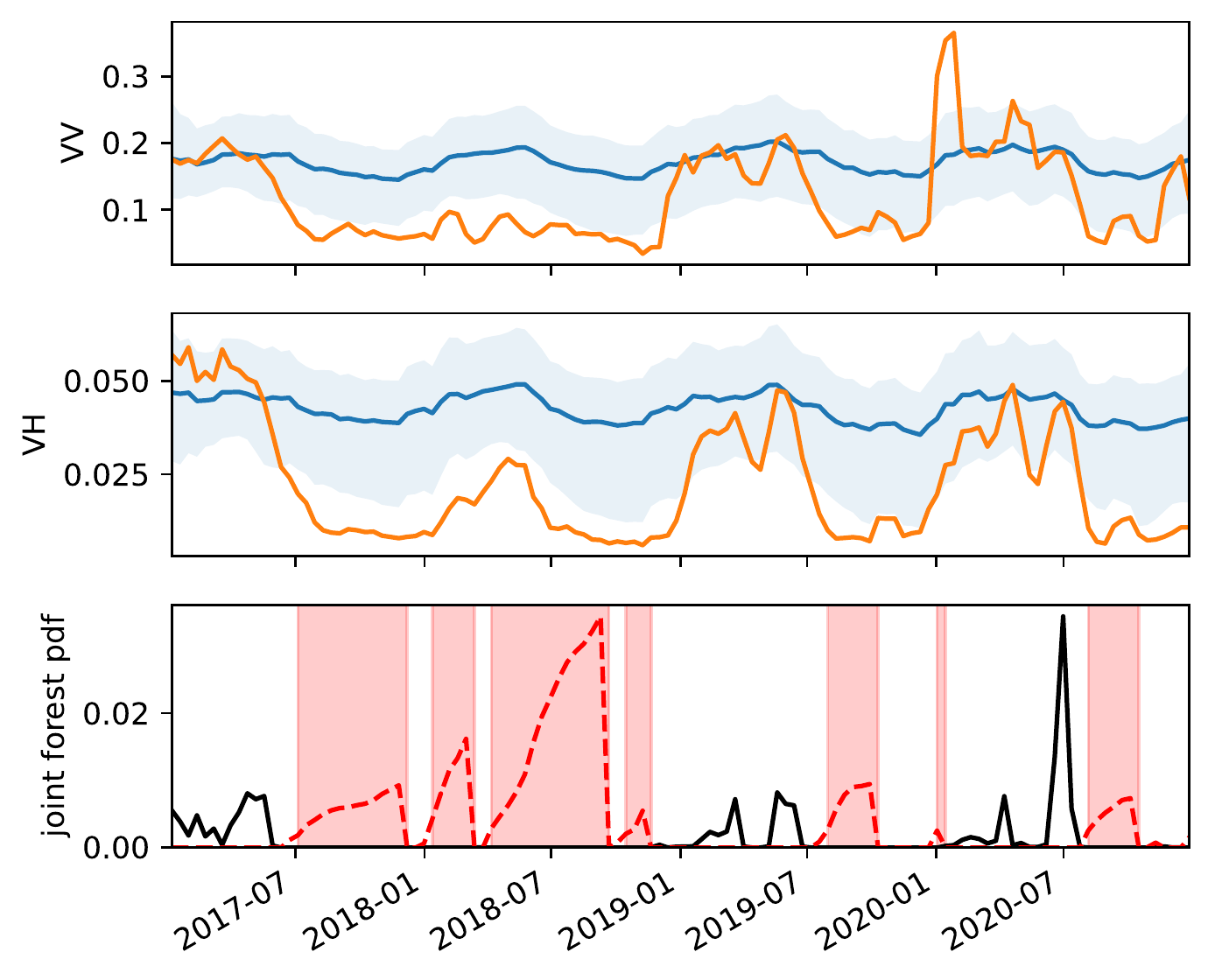}
        \caption{Example 1: Deforestation}
        \label{fig:demo_change_detection_change2}
    \end{subfigure}
    \begin{subfigure}[t]{0.45 \textwidth}
        \includegraphics[width=\textwidth]{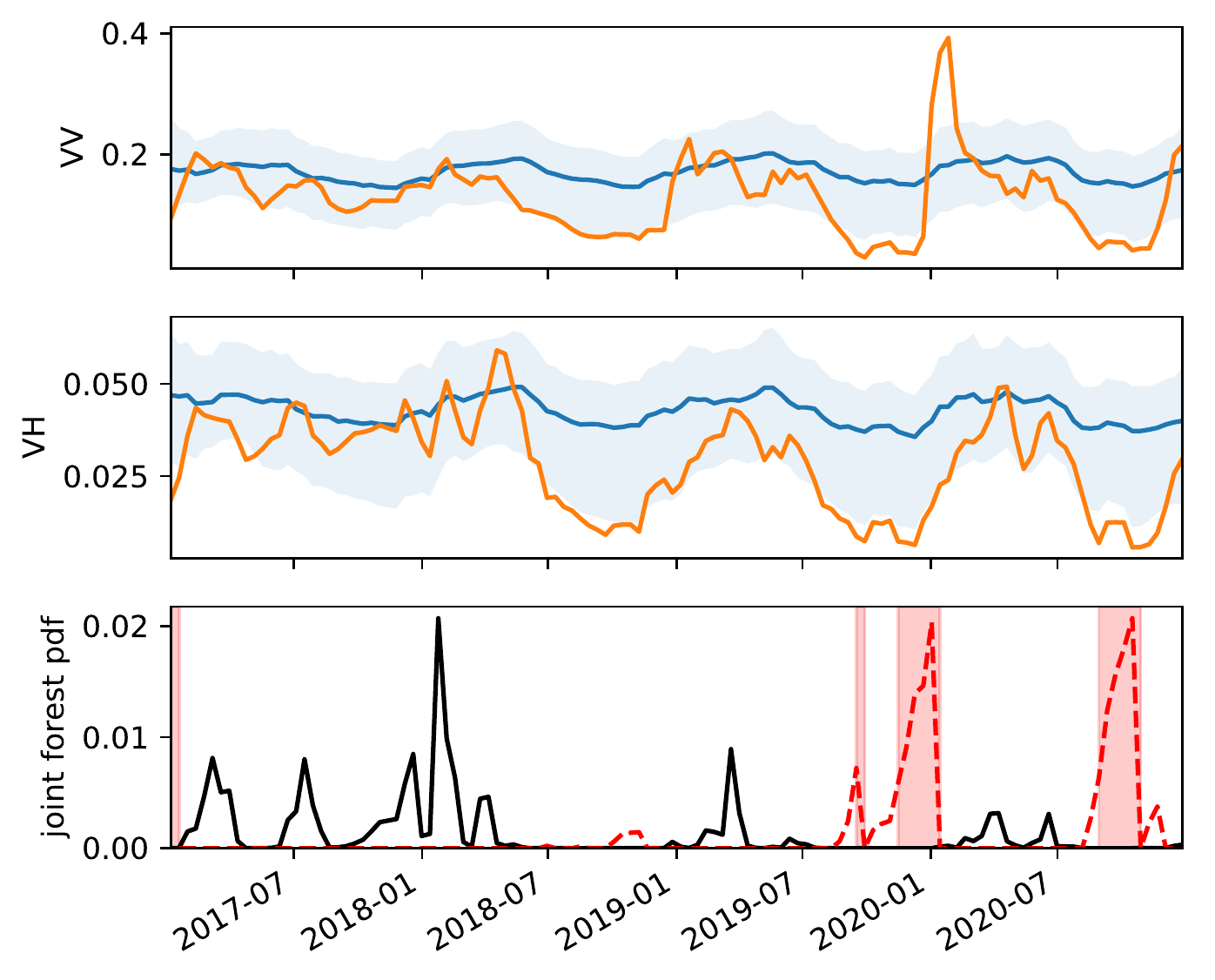}
        \caption{Example 2: Non-forest}
        \label{fig:demo_change_detection_noforest2}
    \end{subfigure}

    \caption{Change detection demonstration for two sample pixels.
    The time series of the pixels for VV and VH is shown in orange,
    the reference forest time series is shown in blue
    (with a light blue band marking the 1\textsuperscript{st}--99\textsuperscript{th} percentile
    within the forest class).
    The third row shows the joint probability density for the pixel being forest at each time step as a black line.
    The dashed red line shows our measure of certainty $\Lambda_t(x)$ for the area being non-forest,
    and the light red areas show such times where the non-forest certainty exceeds a set threshold.
    A deforestation event is then declared at the first encounter of a red area.
    }
    \label{fig:demo_change}
\end{figure}

\Cref{fig:demo_change} shows a demonstration of the change detection method
for two sample time series.
The top two graphs show the time series of the VV amd VH values of each pixel (in orange),
along with the forest mean (in blue).
98\% of all forest pixels fall within the light blue band.
The bottom plots show the forest similarity $p_t(x_t)$ measure in black,
and the cumulative similarity $\Lambda_t(x)$ as a dashed red line.
The areas shaded in red indicate those parts that are considered non-forest based on the threshold $L$.
The first time series (\cref{fig:demo_change_detection_change2}) is of a pixel that undergoes deforestation
early on in the time period.
This event is detected at the start of first shaded area.
The second time series (\cref{fig:demo_change_detection_noforest2}) shows a non-forest pixel
that looks similar to forest for large parts of the time series.

\subsection{Accuracy Assessment} \label{sec:accuracy}
In general,
there are two metrics of interest when assessing the accuracy
of any change map: the \emph{spatial} accuracy
and the \emph{temporal} accuracy.
The spatial accuracy encompasses the precision and recall with
which changes are detected in a binary classification setting,
i.e. the classification of change vs. no change
(independent of the time of change).
The temporal accuracy is a measure of the accuracy of the detected
time of change.
This can be measured as the mean delay of the detected change or the \gls{RMSE}
of the change time.

While the spatial accuracy
can be directly computed from the reference data shown in \cref{fig:reference_change},
the temporal accuracy can only be quantified in a meaningful way
when the precise date of change is known.
The output of the change detection algorithm will always be a precise time of change
that can easily be reduced to a year of change.
However, while comparing the year of change can give some general insight
into the temporal accuracy,
it is not a sufficient measure of the uncertainty in the time of change.
Not only is comparing the year of change a less accurate measure than comparing
e.g. the month of change,
it is in fact qualitatively different:
if the true year of change for one pixel is 2017
but the detected change is for January 2018,
we do not know if this indicates a delay of only one month
---which may be an acceptable detection delay---
or a whole year,
in which case the difference is not merely a delay but likely
the detection of an entirely different change,
i.e. a complete miss of the original first change.
Instead,
we are only going to compute the mean time lag
with respect to the visually interpreted data
as a rough indicator of the temporal accuracy,
while acknowledging that better reference data are necessary for a true
quantification of this uncertainty.

The spatial accuracy is computed by reducing the change result to a binary
map (change / no change).
Several accuracy measures are commonly used for assessing such a map.
All of these can be computed from the entries of a confusion matrix,
i.e., the number of true positives (TP), true negatives (TN),
false positives (FP), and false negatives (FN).
The \gls{PA} (also known as sensitivity, recall, or true positive rate)
in this context is the probability that an actual change is accurately detected.
It is defined as
\begin{equation} \label{eq:producers_acc}
    \PA = \frac{\TP}{\TP + \FN}
\end{equation}

The \gls{UA} (also known as precision or positive predictive value)
is defined as the probability that a change shown on the map corresponds to a true change on the ground.
It is given by
\begin{equation} \label{eq:users_acc}
    \UA = \frac{\TP}{\TP + \FP}
\end{equation}

The \gls{OA} (or simply accuracy) is defined as
\begin{equation} \label{eq:overall_acc}
    \OA = \frac{\TP + \TN}{\TP + \FP + \TN + \FN}
\end{equation}
and is the overall fraction of correctly classified pixels as change or no change.
The \gls{OA} can be a misleading accuracy measure as it depends on
the fraction of pixels that that are subject to change.
A more robust measure is the \gls{BA}
which accounts for class imbalances,
i.e., the case where the available classes
have different numbers of members \citep{Brodersen2010}.
It is defined as
\begin{equation}\label{eq:balanced_accuracy}
\BA = \frac{1}{2} \left( \frac{\TP}{\TP+\FP} + \frac{\TN}{\TN+\FN} \right).
\end{equation}

\FloatBarrier
\section{Results}

\begin{figure}[!h]
    \centering
    \begin{subfigure}[c]{0.32 \textwidth}
        \includegraphics[width=\textwidth]{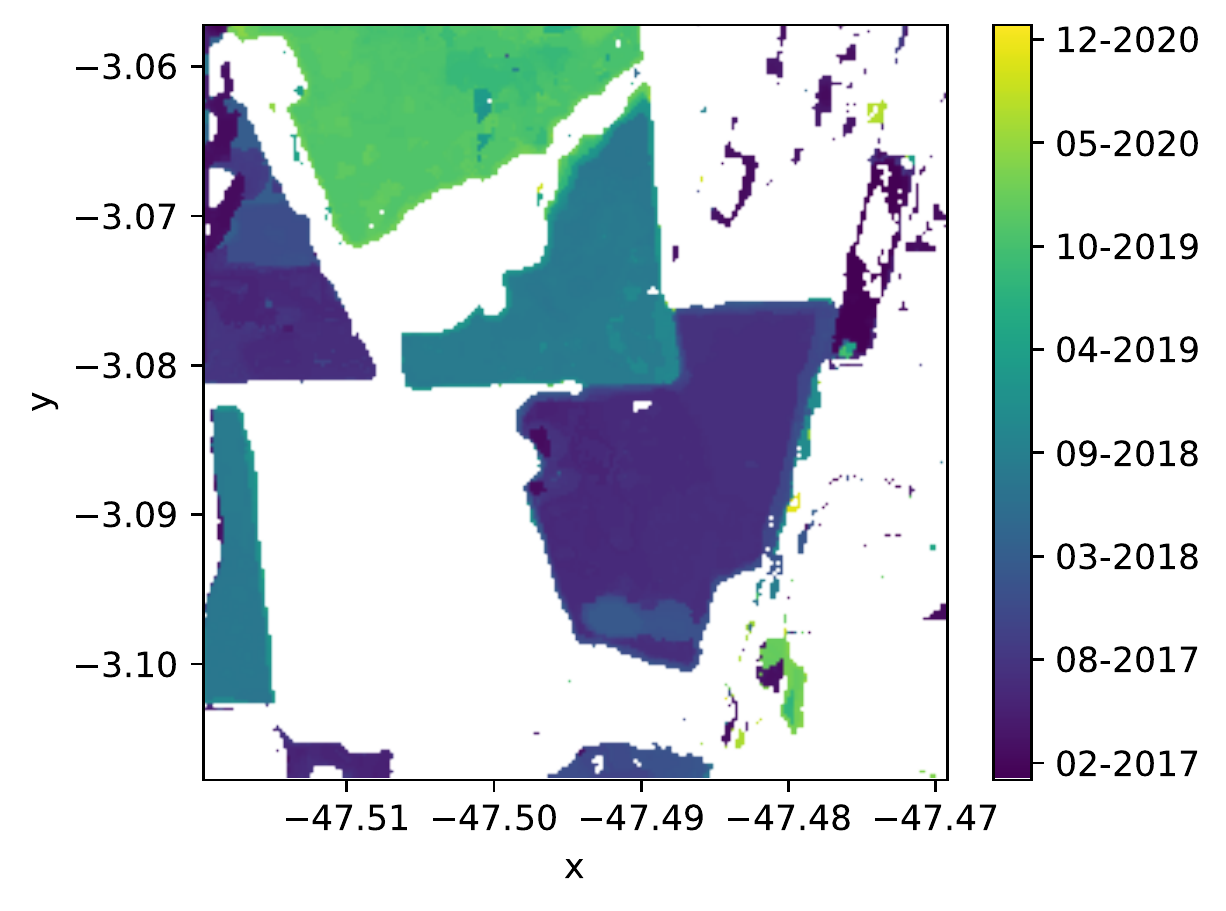}
        \caption{Change detection map}
        \label{fig:changemap_paragominas}
    \end{subfigure}
    \begin{subfigure}[c]{0.32 \textwidth}
        \includegraphics[width=\textwidth]{cmaps/reference_polygons.pdf}
        \caption{Reference map}
        \label{fig:reference_polygons}
    \end{subfigure}
    \begin{subfigure}[c]{0.32 \textwidth}
        \includegraphics[width=\textwidth]{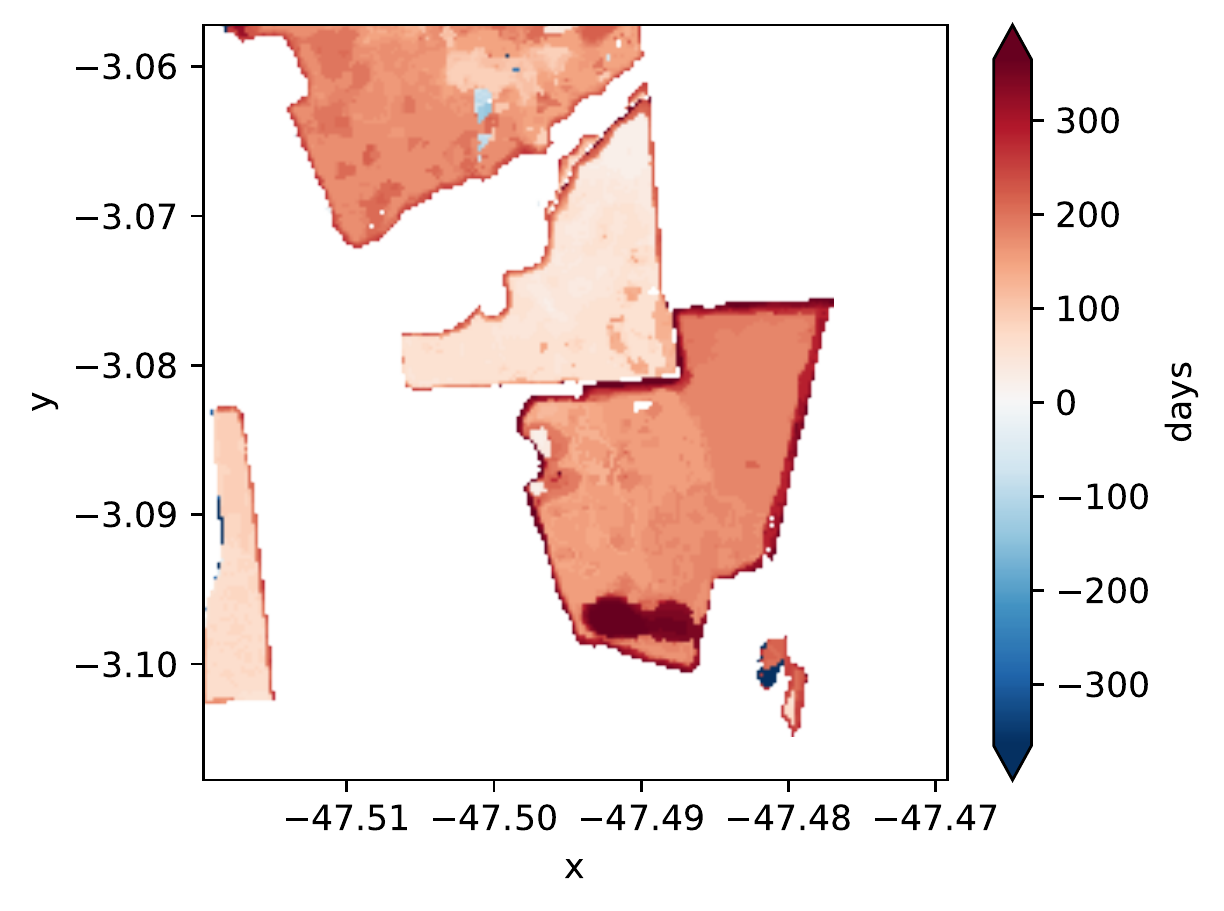}
        \caption{Change detection delay}
        \label{fig:changemap_paragominas_lag}
    \end{subfigure}

    \caption{This figure shows a comparison of the detected date of change (\cref{fig:changemap_paragominas})
    and the reference map obtained from manual interpretation (\cref{fig:reference_polygons}).
    For those pixels where both maps indicate a change,
    \cref{fig:changemap_paragominas_lag} shows the change detection delay in days.
    }
    \label{fig:change_result_paragominas}
\end{figure}

\Cref{fig:change_result_paragominas} shows the result of the algorithm
for the site in Paragominas, Brazil (\cref{fig:demo_distribution}) for a time series from 2017--2020.
The change detection delay in days is shown in \cref{fig:changemap_paragominas_lag}.
This is computed with respect to the true deforestation dates as extracted via visual inspection
of Landsat and Planet data (\cref{fig:reference_polygons}).

A comparison of the resulting change map (\cref{fig:changemap_paragominas})
with the reference map from visual interpretation (\cref{fig:reference_polygons})
shows that all areas of deforestation are picked up reliably.
In addition, however,
deforestation is falsely detected in several areas where in fact no deforestation is happening. All of these are areas that have undergone deforestation in the past.
For example, the patch in the north-western corner of the area is being picked up as deforestation,
when in fact it was non-forest even at the start of the time series.
The backscatter signature of these areas looks very similar to that of forest
at the start of the time series,
such that the algorithm is confused into detecting a transition from forest to non-forest
when in reality the pixel was non-forest throughout the time period.
The achieved \gls{PA} is 96.5\%, with a \gls{UA} of 75.7\% and a balanced accuracy of 90.4\%,
reflecting these findings.

\begin{figure}[!p]
    \centering
    \begin{subfigure}[t]{\textwidth}
        \includegraphics[width=\textwidth]{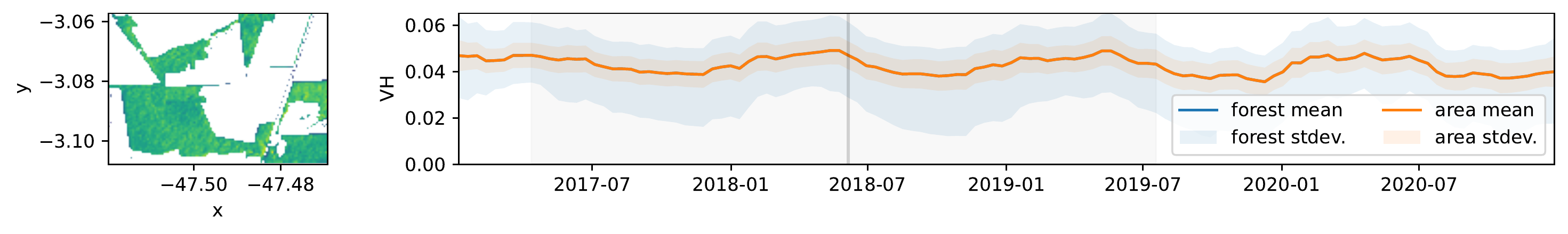}
    \end{subfigure}
    \begin{subfigure}[t]{\textwidth}
        \includegraphics[width=\textwidth]{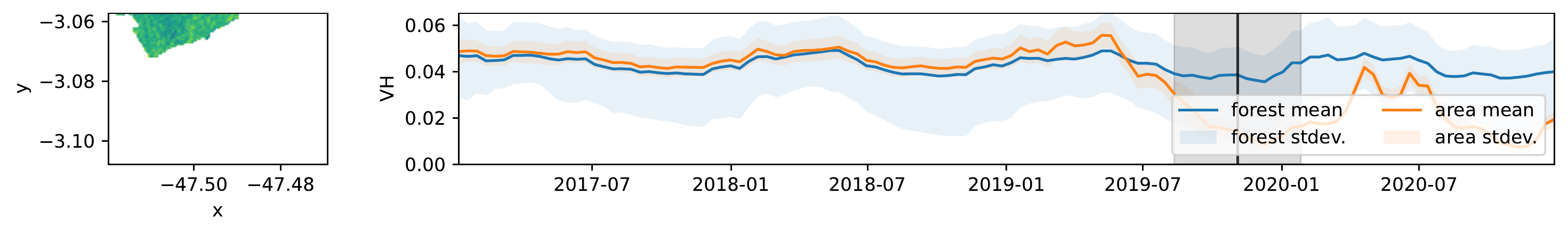}
    \end{subfigure}
    \begin{subfigure}[t]{\textwidth}
        \includegraphics[width=\textwidth]{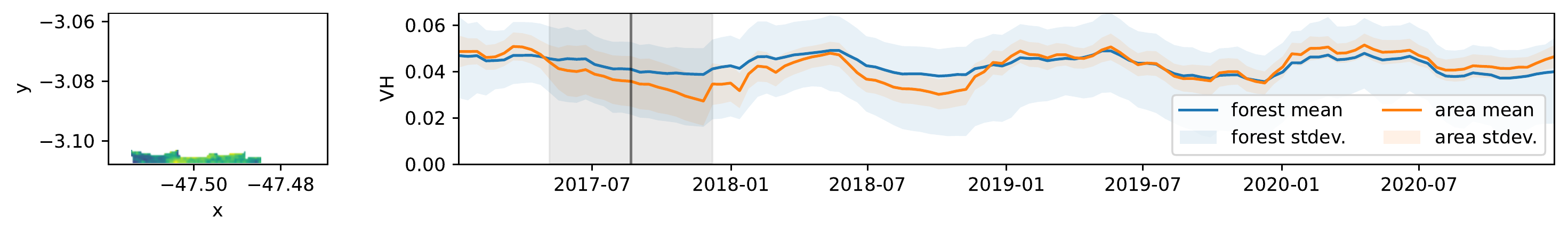}
    \end{subfigure}
    \begin{subfigure}[t]{\textwidth}
        \includegraphics[width=\textwidth]{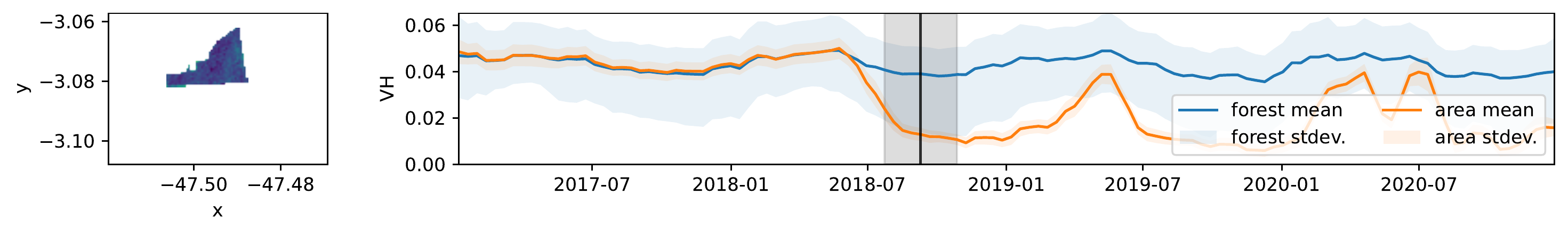}
    \end{subfigure}
    \begin{subfigure}[t]{\textwidth}
        \includegraphics[width=\textwidth]{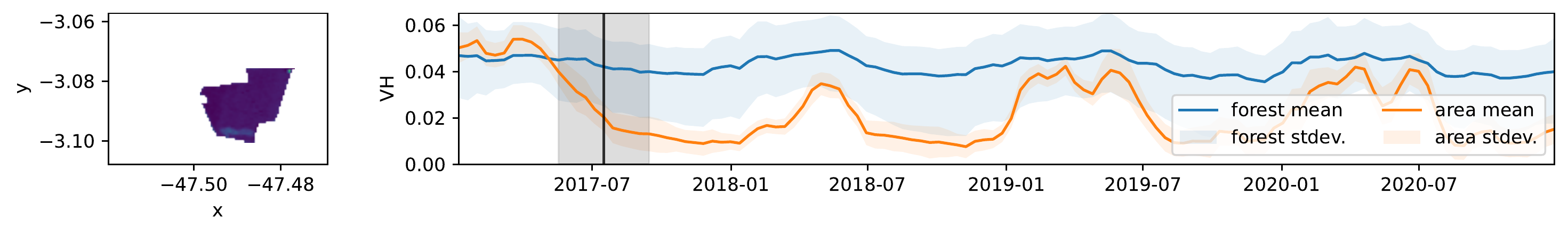}
    \end{subfigure}
    \begin{subfigure}[t]{\textwidth}
        \includegraphics[width=\textwidth]{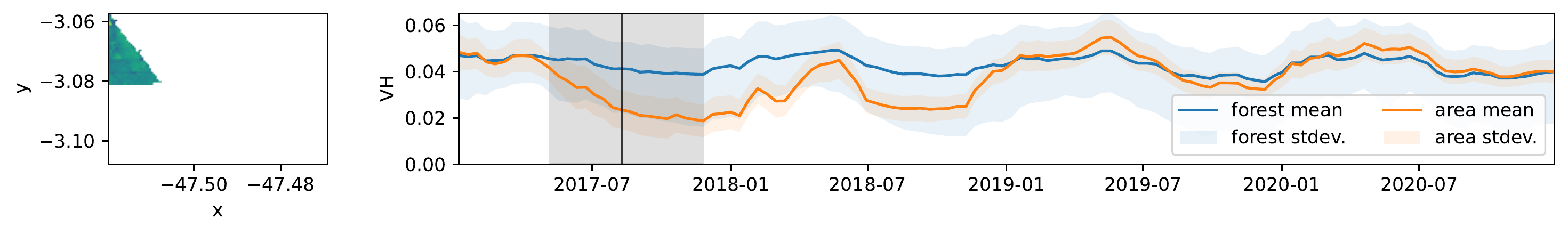}
    \end{subfigure}
    \begin{subfigure}[t]{\textwidth}
        \includegraphics[width=\textwidth]{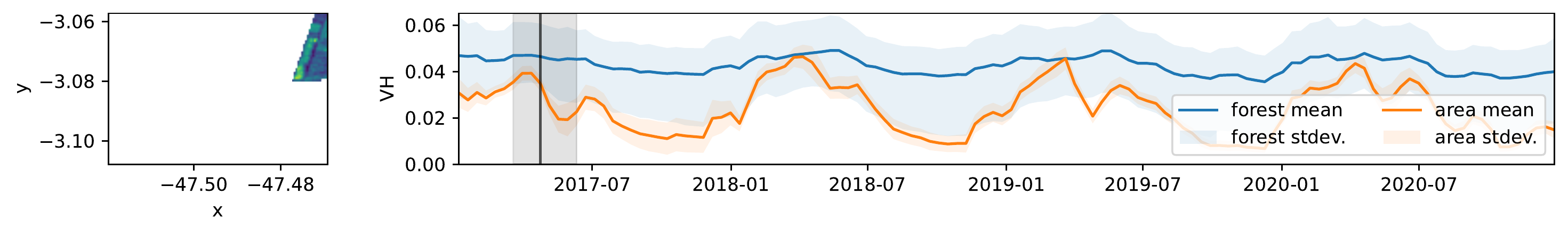}
    \end{subfigure}
    \begin{subfigure}[t]{\textwidth}
        \includegraphics[width=\textwidth]{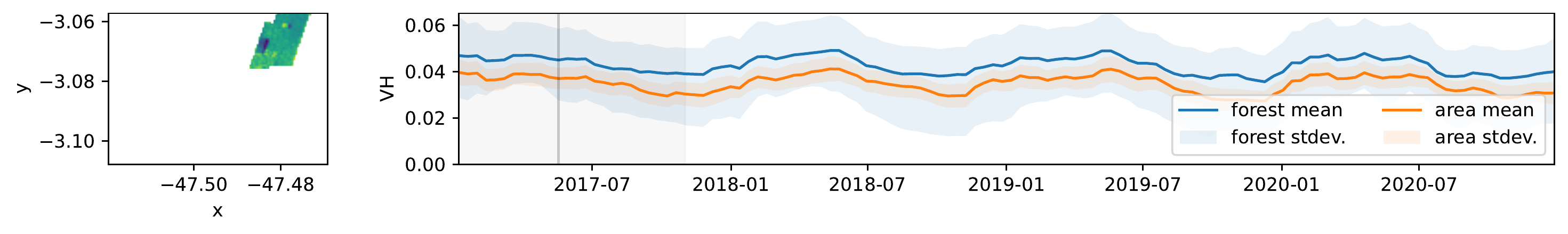}
    \end{subfigure}
    \caption{This figure shows the mean time series for some of the polygons
    shown in \cref{fig:demo_mask}.
    It also shows the mean change point detected for this area
    as a vertical gray line.
    The change point variation in the area is shown as a gray band around this line.
    The opacity of the change point line indicates the fraction of detected change points --
    for the first and last row, for example,
    change points were only detected for a small number of pixels.
    The blue line and light blue band indicate the reference forest time series mean
    and 1\textsuperscript{st}--99\textsuperscript{th} percentile,
    respectively.
    }
    \label{fig:paragominas_areas}
\end{figure}

\Cref{fig:paragominas_areas} shows the mean time series for some of the polygons
delineated in \cref{fig:reference_visual}.
The solid blue line indicates the forest reference time series,
with the light blue band marking the 1\textsuperscript{st}--99\textsuperscript{th} percentile
of forest pixels.
In addition, the mean change point is indicated for each polygon,
along with the standard deviation (marked as vertical gray band around the change point).
The opacity of the marked change point and its uncertainty is scaled
with the fraction of pixels in the area for which a change point was detected.

The first row represents stable forest.
While changes are detected for some pixels in this area,
e.g. due to imperfect polygon boundaries,
the algorithm only produces very few false positives over forested areas,
as visually indicated by the low opacity of the marked mean change point.
On the other hand, the third row is an examples of an area  
that is non-forest throughout the time period,
but is consistently classified as deforestation events using the method shown.
The third row in particular, and also the sixth and last row,
are non-forest but look very similar to forest in their temporal radar signature,
confusing the algorithm into mistakenly predicting deforestation
--- illustrating the reason for the false positives detected in the change map above.

\FloatBarrier
\subsection{Validation at additional sites}

\begin{figure}[!htb]
    \centering
    \begin{subfigure}[c]{0.3 \textwidth}
        \includegraphics[width=\textwidth]{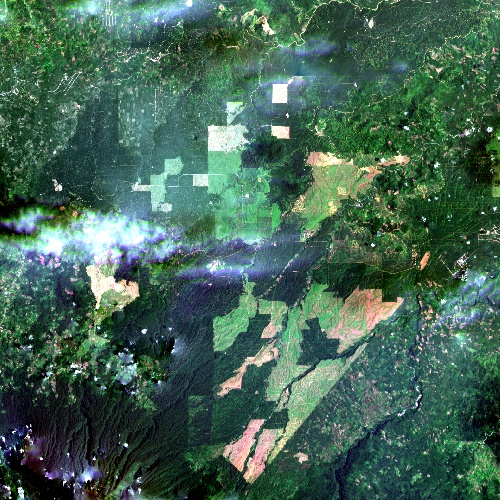}
        \caption{}
        \label{fig:nicfi_riau}
    \end{subfigure}
    \begin{subfigure}[c]{0.3 \textwidth}
        \includegraphics[width=\textwidth]{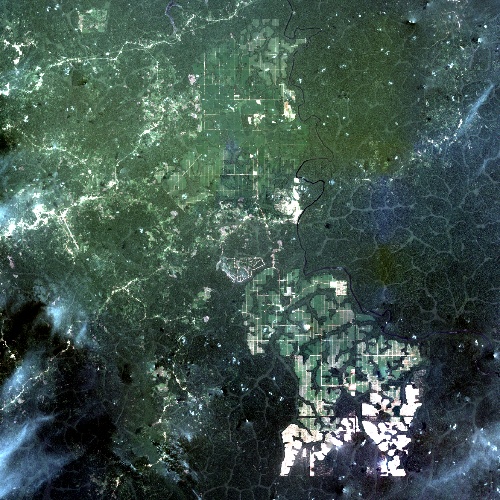}
        \caption{}
        \label{fig:nicfi_cameroon}
    \end{subfigure}
    \begin{subfigure}[c]{0.3 \textwidth}
        \includegraphics[width=\textwidth]{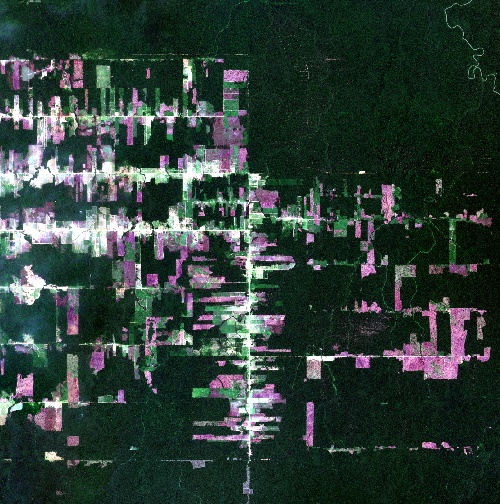}
        \caption{}
        \label{fig:nicfi_brazil}
    \end{subfigure}

    \begin{subfigure}[c]{0.3 \textwidth}
        \includegraphics[width=\textwidth]{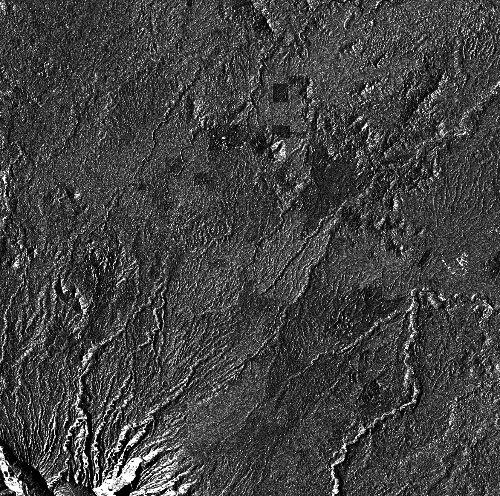}
        \caption{}
        \label{fig:s1_riau}
    \end{subfigure}
    \begin{subfigure}[c]{0.3 \textwidth}
        \includegraphics[width=\textwidth]{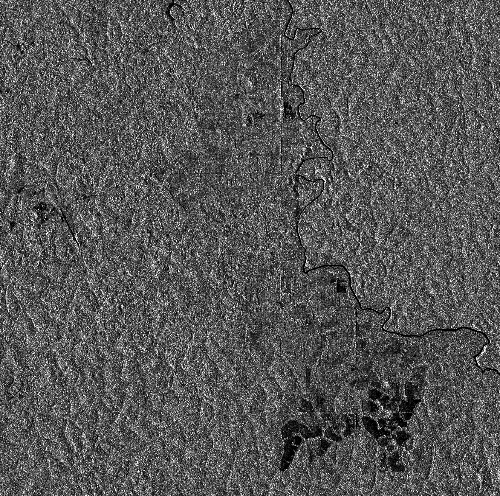}
        \caption{}
        \label{fig:s1_cameroon}
    \end{subfigure}
    \begin{subfigure}[c]{0.3 \textwidth}
        \includegraphics[width=\textwidth]{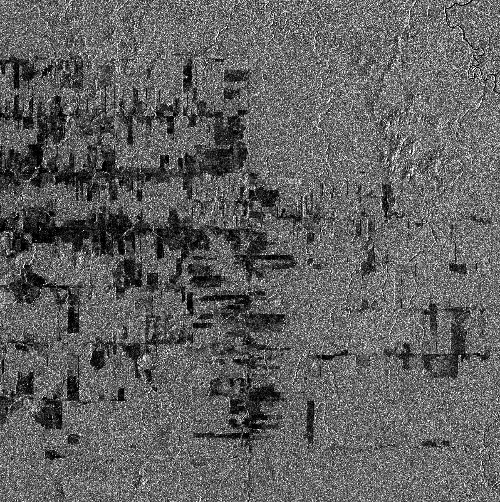}
        \caption{}
        \label{fig:s1_brazil}
    \end{subfigure}

    \caption{This figure shows individual Sentinel-1 images (VV)
    and NICFI images of the same time period for comparison,
    for each of the additional validation sites.
    The images shown are for Riau, Indonesia (October 2021, \cref{fig:nicfi_riau,fig:s1_riau}),
    Cameroon (December 2018, \cref{fig:nicfi_cameroon,fig:s1_cameroon}),
    and Porto Velho (\cref{fig:nicfi_brazil,fig:s1_brazil}).}
    \label{fig:nicfi_s1_validation}
\end{figure}

To validate that the method works in diverse biomes and conditions,
it was tested on three addition sites with distinct deforestation and vegetation change patterns.
The sites chosen are located in
(1) Porto Velho, Brazil (2017--2021),
(2) South Cameroon (2017--2018),
and (3) Riau, Indonesia (2020--2021).
\Cref{fig:nicfi_s1_validation} shows an individual Sentinel-1 image
as well as a NICFI \citep{NICFI} RGB composite for each of the sites.

\begin{figure}[!h]
    \centering
    \begin{subfigure}[t]{0.3 \textwidth}
        \includegraphics[width=\textwidth]{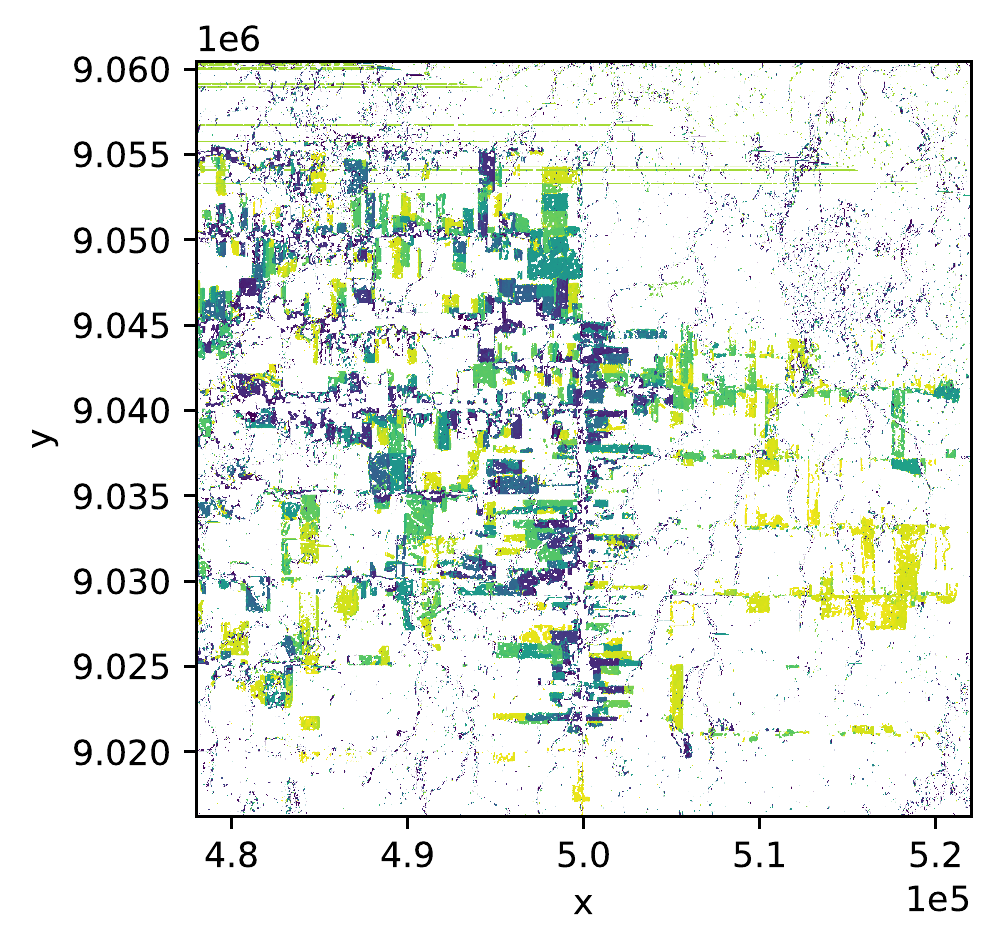}
        \caption{Porto Velho Changes 2017--2021}
        \label{fig:brazil_cmap}
    \end{subfigure}
    \begin{subfigure}[t]{0.3 \textwidth}
        \includegraphics[width=\textwidth]{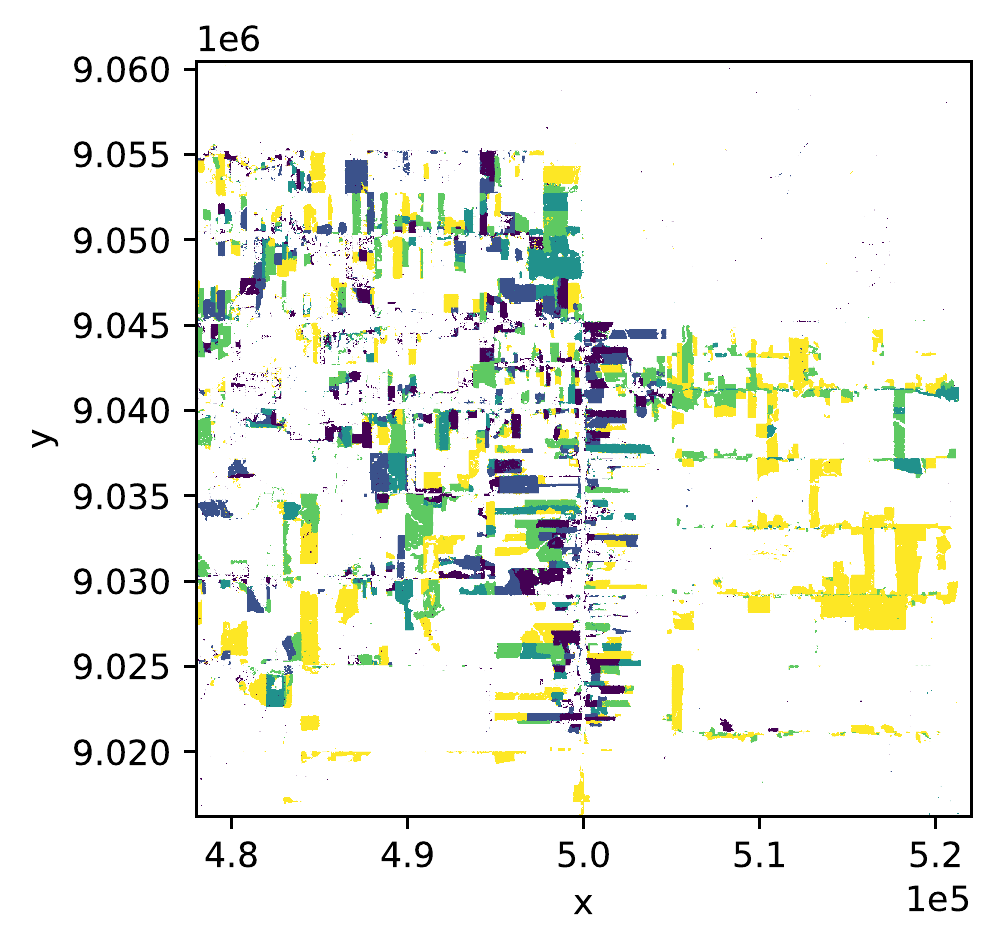}
        \caption{Hansen Porto Velho 2017--2021}
        \label{fig:brazil_hansen}
    \end{subfigure}
    \begin{subfigure}[t]{0.3 \textwidth}
        \includegraphics[width=\textwidth]{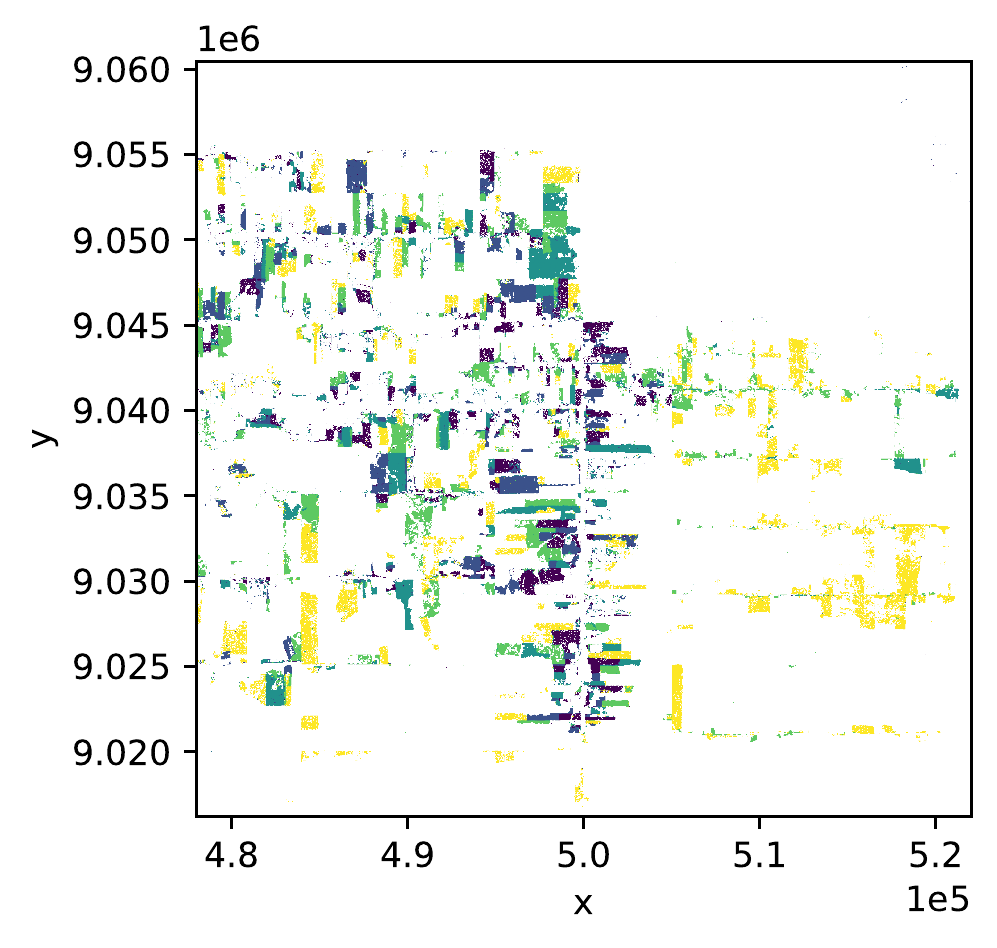}
        \caption{JRC TMF Porto Velho 2017--2021}
        \label{fig:brazil_jrc}
    \end{subfigure}
    \begin{subfigure}[t]{0.08 \textwidth}
        \includegraphics[width=\textwidth, trim=0 -1cm 0 0]{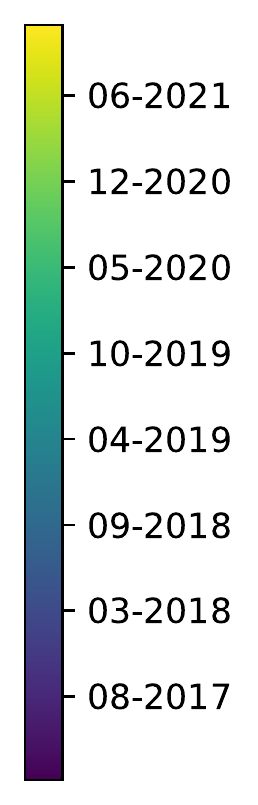}
    \end{subfigure}

    \begin{subfigure}[t]{0.3 \textwidth}
        \includegraphics[width=\textwidth]{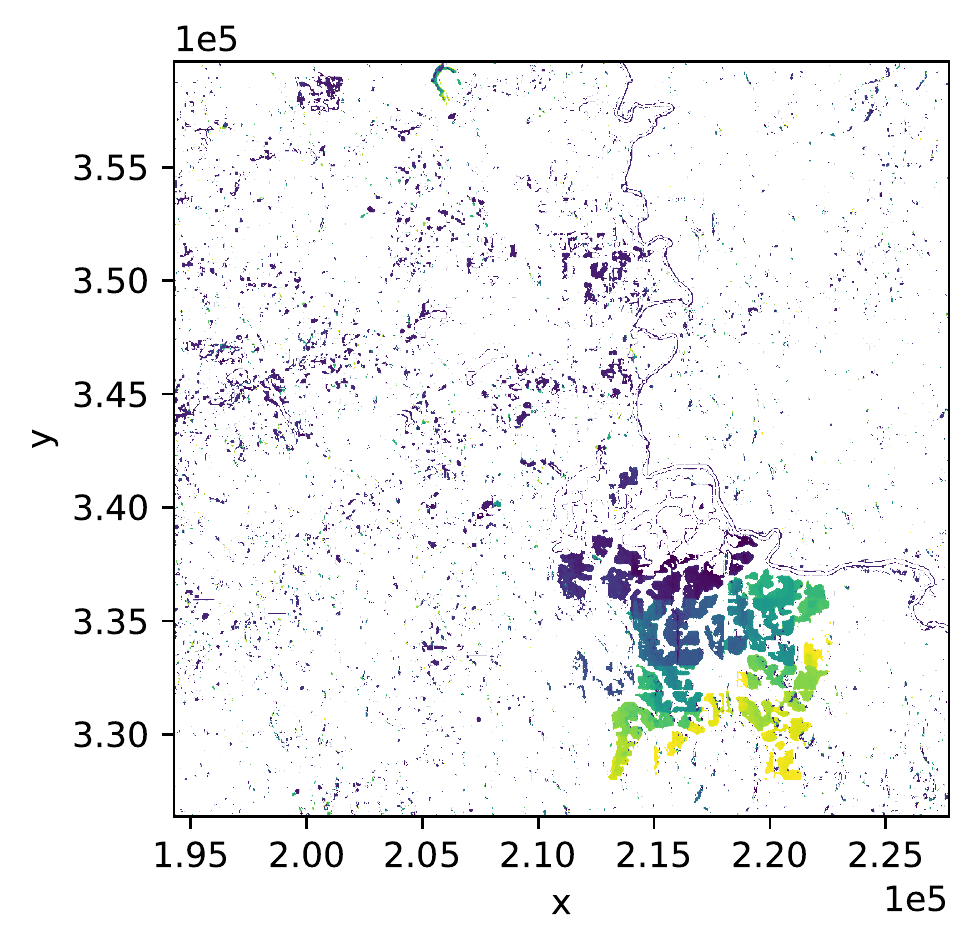}
        \caption{Cameroon Changes 2017--2018}
        \label{fig:cameroon_cmap}
    \end{subfigure}
    \begin{subfigure}[t]{0.3 \textwidth}
        \includegraphics[width=\textwidth]{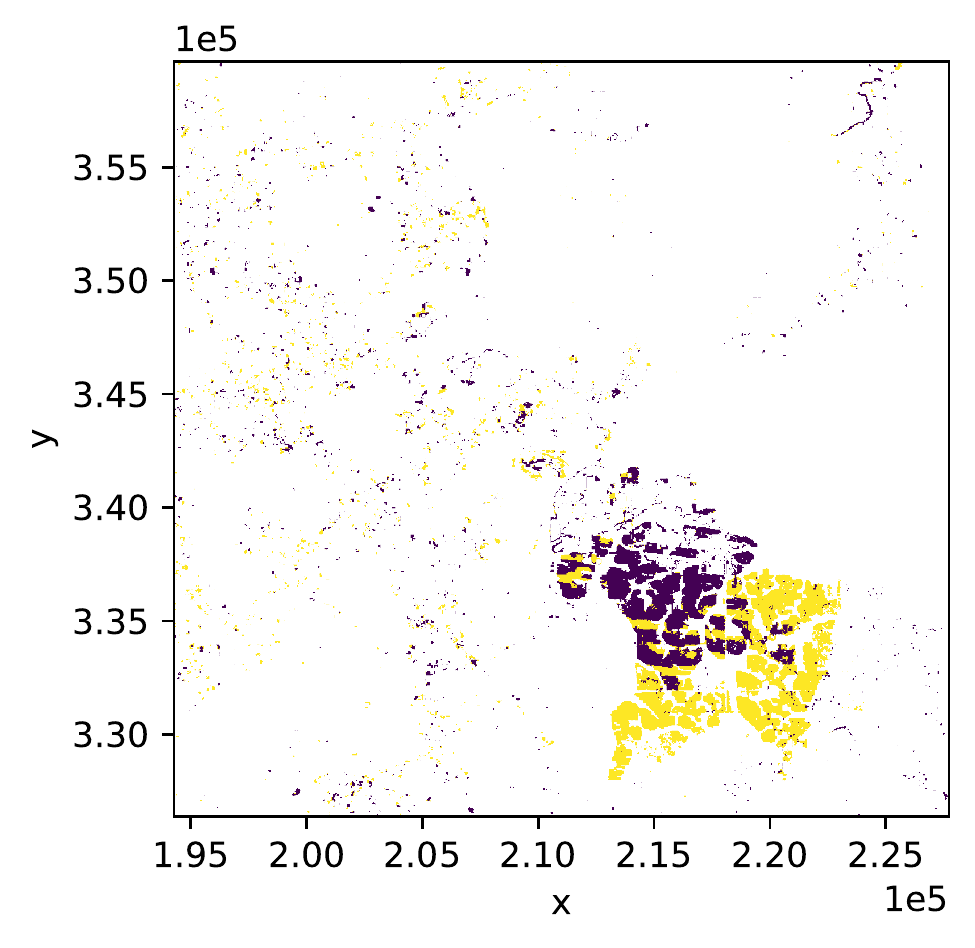}
        \caption{Hansen Cameroon 2017--2018}
        \label{fig:cameroon_hansen}
    \end{subfigure}
    \begin{subfigure}[t]{0.3 \textwidth}
        \includegraphics[width=\textwidth]{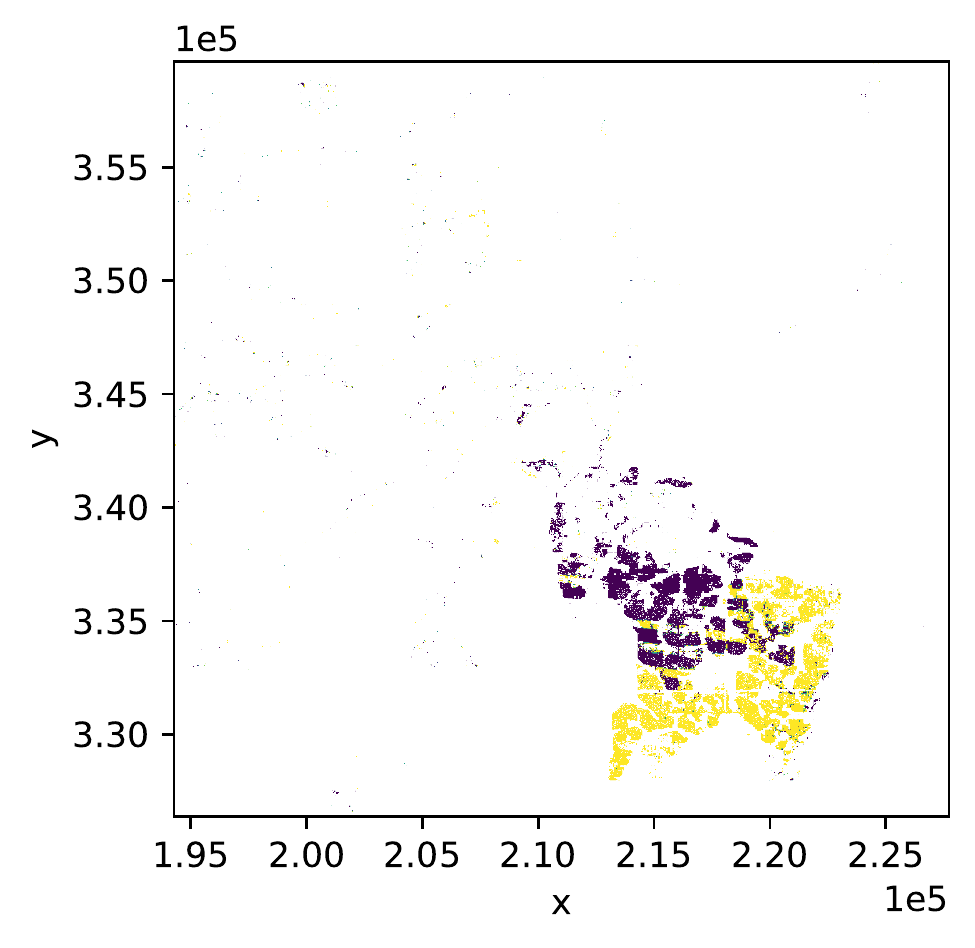}
        \caption{JRC TMF Cameroon 2017--2018}
        \label{fig:cameroon_jrc}
    \end{subfigure}
    \begin{subfigure}[t]{0.08 \textwidth}
        \includegraphics[width=\textwidth, trim=0 -1cm 0 0]{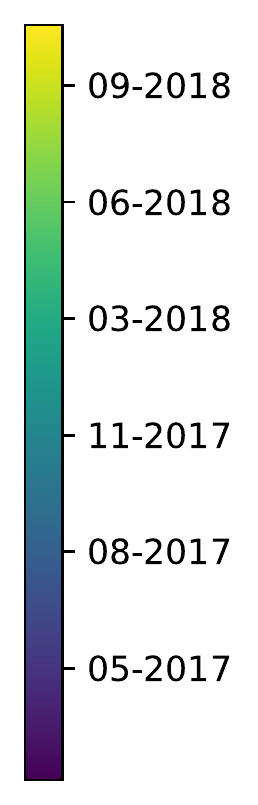}
    \end{subfigure}

    \begin{subfigure}[t]{0.3 \textwidth}
        \includegraphics[width=\textwidth]{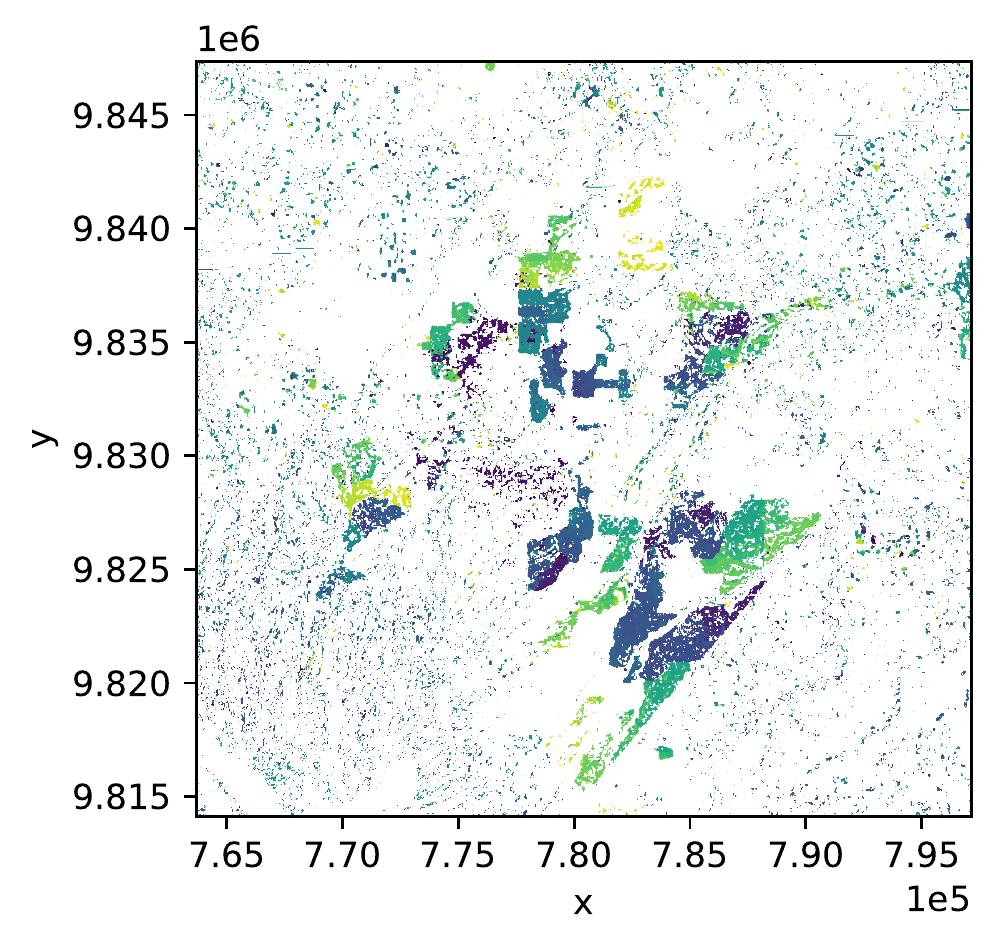}
        \caption{Riau Changes 2020--2021}
        \label{fig:riau_cmap}
    \end{subfigure}
    \begin{subfigure}[t]{0.3 \textwidth}
        \includegraphics[width=\textwidth]{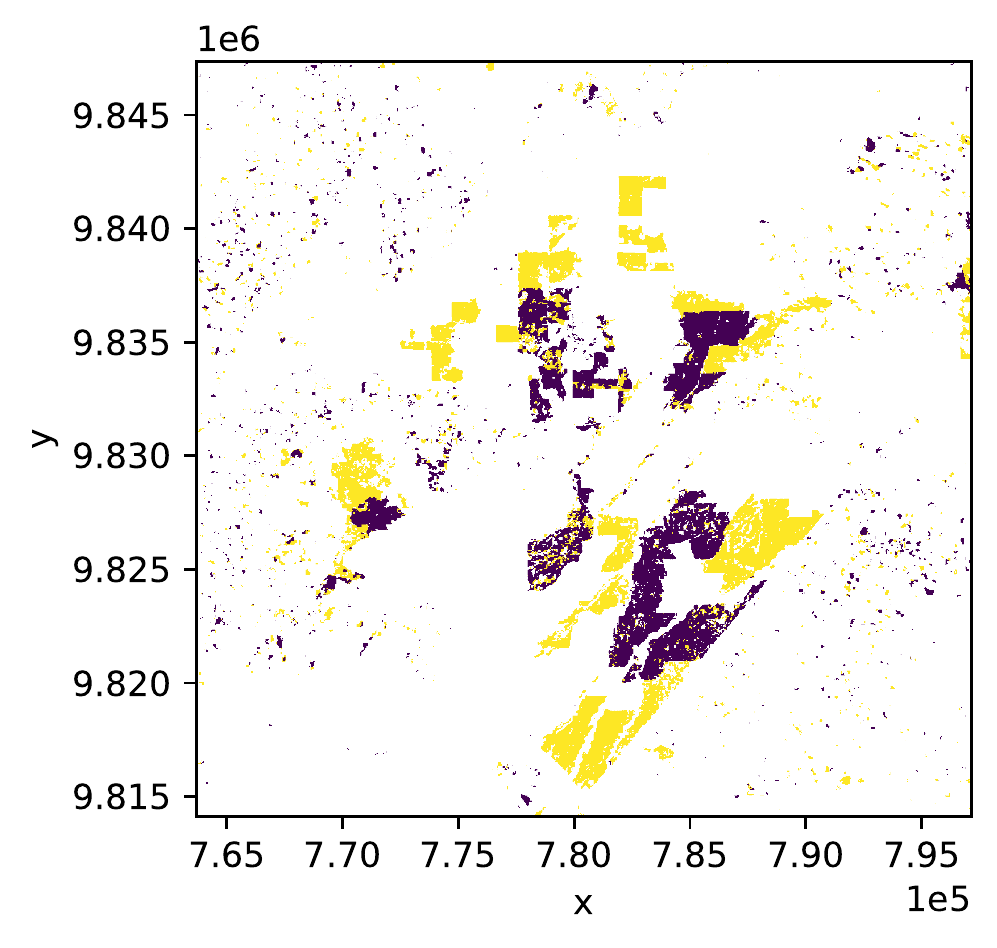}
        \caption{Hansen Riau 2020--2021}
        \label{fig:riau_hansen}
    \end{subfigure}
    \begin{subfigure}[t]{0.3 \textwidth}
        \includegraphics[width=\textwidth]{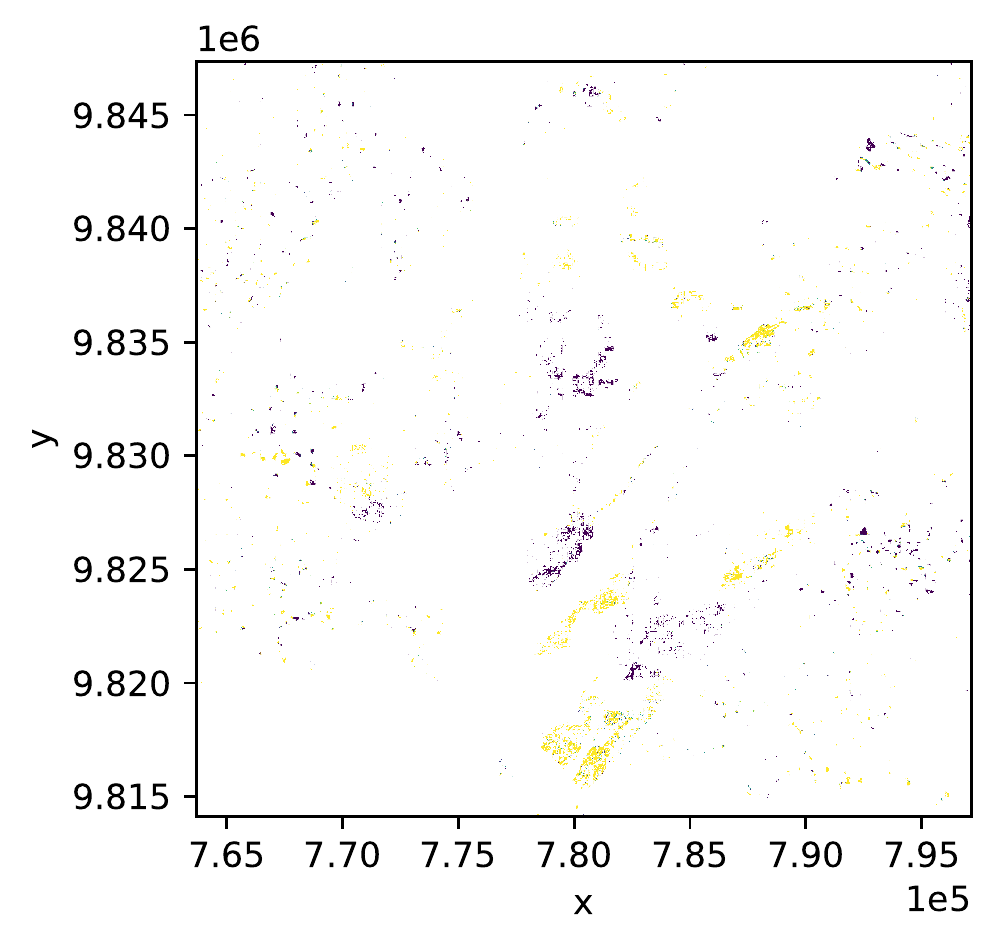}
        \caption{JRC TMF Riau 2020--2021}
        \label{fig:riau_jrc}
    \end{subfigure}
    \begin{subfigure}[t]{0.08 \textwidth}
        \includegraphics[width=\textwidth, trim=0 -1cm 0 0]{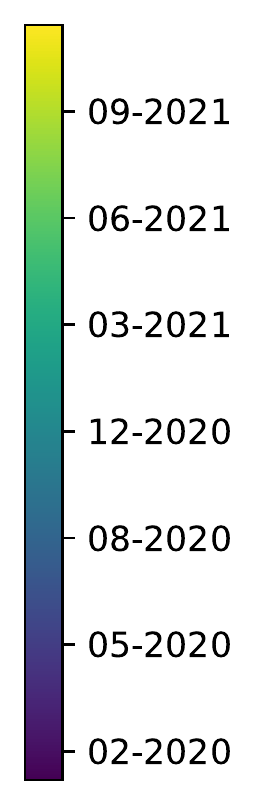}  
    \end{subfigure}

    \caption{This figure shows the deforestation maps obtained at three additional validation sites,
    along with the reference maps from \citet{Hansen2013} and \citet{Vancutsem2021}.}
    \label{fig:cmaps}
\end{figure}

It is immediately evident that the deforestation is harder to detect with Sentinel-1
than with optical sensors.
While deforestation can be clearly observed in Brazil and partially also in Cameroon,
the deforested patches in Indonesia are almost invisible to the eye in the Sentinel-1 image.
Detection is further complicated by the presence of mountainous regions in the south western corner.
The resulting change detection output is shown in \cref{fig:cmaps}.
The maps on the left show the change detection results,
the middle maps are the Hansen reference \citep{Hansen2013},
and the maps on the right are the JRC TMF deforestation maps \citep{Vancutsem2021}.
For these sites,
no deforestation reference was extracted by visual intepretation.
For that reason, only the year of change is known from the two reference maps.
The reference maps therefore show discrete change years,
while the change detection map on the left shows a continuous change date
as indicated by the more graduated color scheme.


In addition to being able to provide an actual date of change for each deforestation event,
it is apparent from the Cameroon maps that the Sentinel-1 based change maps are not affected
by the stripy Landsat artifacts that are apparent in both the Hansen and JRC maps.
However,
there are also limitations that are
evident in the resulting change maps.
For example,
the change maps appear to be generally noisier than the reference maps.
This noise may be partially due to false positives in non-forest areas as
discussed previously,
but may also be an artifact of the Sentinel-1 data used.
While Sentinel-1 data can definitely separate forest and non-forest
when aggregated annually,
separation at an individual time step is unreliable
because of speckle and seasonality \citep{Hansen2020}.
Optical data are generally better suited for differentiating these classes
at a single time step,
assuming a cloud-free image.
Therefore, a direct comparison between a day-by-day change map from SAR data
and an annual map from optical data
in terms of spatial noise is an unfair comparison
as the annual map can be considered to have applied a large amount of temporal smoothing
to eliminate noise and at the same time get around the problem of missing data.

\Cref{tab:accuracy} summarizes the accuracy metrics computed for each site
with respect to the various available reference maps.
It shows the \gls{UA}, \gls{PA}, and \gls{BA} (see~\cref{sec:accuracy}).
The thresholds $q$ and $L$ were chosen to optimize the BA with respect to the Hansen maps.

\begin{table}[!htb]
    \renewcommand{\arraystretch}{1.5}
    \centering
    \caption{This table summarizes the accuracy metrics \gls{UA}, \gls{PA}, and \gls{BA}
    for each site,
    with respect to the Hansen and JRC maps,
    as well as a visual interpretation in the case of the Paragominas site.}
    \begin{tabular}{llrrrrr}
        \toprule
        \textbf{Reference} & \textbf{Site} & \textbf{UA} & \textbf{PA} & \textbf{BA}\\
        \midrule
        \multirow{4}{*}{Hansen} & Paragominas & \scorecolor{77.7} & \scorecolor{86.5} & \scorecolor{85.5} \\
        & Cameroon & \scorecolor{48.0} & \scorecolor{67.7} & \scorecolor{81.4} \\
        & Riau & \scorecolor{49.8} & \scorecolor{61.0} & \scorecolor{76.9} \\
        & Porto Velho & \scorecolor{62.0} & \scorecolor{76.4} & \scorecolor{83.2} \\

        \midrule
        \multirow{4}{*}{JRC} & Paragominas & \scorecolor{57.9} & \scorecolor{91.7} & \scorecolor{83.6} \\
        & Cameroon & \scorecolor{34.9} & \scorecolor{82.2} & \scorecolor{88.1} \\
        & Riau & \scorecolor{5.3} & \scorecolor{48.3} & \scorecolor{68.0} \\
        & Porto Velho & \scorecolor{42.9} & \scorecolor{84.0} & \scorecolor{85.0} \\

        \midrule
        \multirow{1}{*}{Visual} & Paragominas & \scorecolor{75.7} & \scorecolor{96.5} & \scorecolor{90.4} \\

        \bottomrule


    \end{tabular}
    \label{tab:accuracy}
\end{table}

The accuracy scores show some variance between the sites.
It is evident that the method generally achieves a high \gls{PA},
potentially at the cost of a lower \gls{UA}.
The lower \gls{UA} is driven by the false positives observed.
The most reliable accuracy scores are those based on the visual interpretation,
as the Hansen and JRC maps carry errors in themselves and do not necessarily reflect
the ground truth.
This is particularly evident for the JRC based accuracy scores for the Indonesian site.
These scores are very low as the JRC map fails to detect any deforestation in this area.
These issues serve to illustrate the fact that the accuracy scores do not fully reflect
the performance of the change detection method.
The method may in fact outperform the Hansen and JRC maps,
in which case a lower score would reflect negatively on these reference maps instead.
However, they do serve as a useful sanity check of the results.

Indonesia is a challenging site as the forested areas look very similar to the non-forest
areas in the backscatter (see \cref{fig:nicfi_s1_validation}).
It also exhibits more terrain than the other two sites,
further complicating the use of \gls{SAR} data.
In general,
balanced accuracy scores of about 80\% were observed
with respect to both the Hansen and JRC maps.
The \gls{PA} and \gls{UA} scores vary strongly between sites.
The \gls{UA} is generally low due to the number of false positives.
The \gls{PA} is as low as 60--70\% for Indonesia and Cameroon (compared to Hansen),
and as high as 80--90\% for the two Brazilian sites.


\FloatBarrier
\subsection{Noisy reference labels}

Given that the idea behind this method is to be relatively
independent of high quality training data,
in this section the behavior of the method in the presence of mislabeled data
is investigated.
To simulate corrupted training data,
a forest reference ensemble of size $N$ is generated for
a range of fractions $c$ of incorrectly classified forest pixels
such that $(1-c)N$ ensemble members are true forest pixels,
and $cN$ ensemble members are in fact non-forest.
For each such corrupted reference ensemble,
a change map was generated and the corresponding accuracy values were computed.

\begin{figure}[!ht]
    \centering
    \begin{subfigure}[t]{0.15 \textwidth}
        \includegraphics[width=\textwidth]{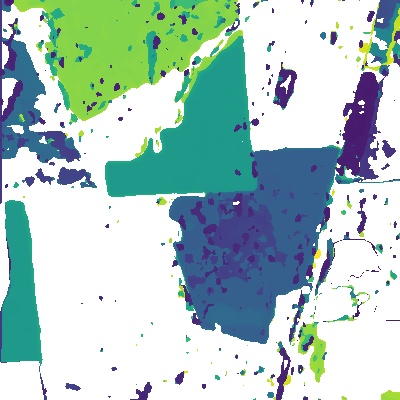}
        \caption*{0\%}
        \label{fig:cmap_noise_0.00}
    \end{subfigure}
    \begin{subfigure}[t]{0.15 \textwidth}
        \includegraphics[width=\textwidth]{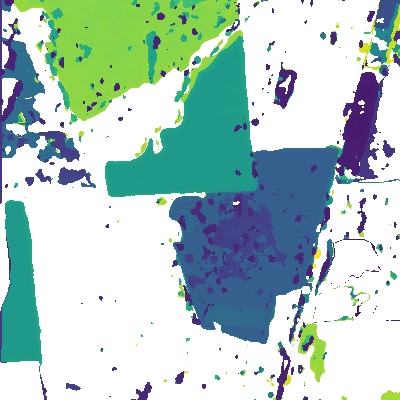}
        \caption*{2\%}
        \label{fig:cmap_noise_0.02}
    \end{subfigure}
    \begin{subfigure}[t]{0.15 \textwidth}
        \includegraphics[width=\textwidth]{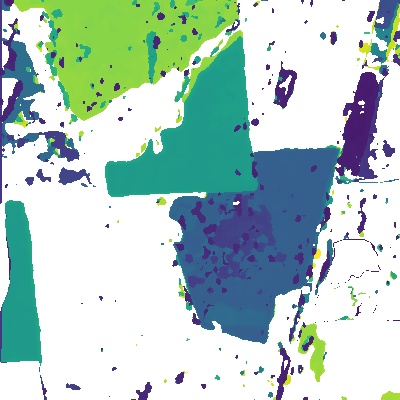}
        \caption*{4\%}
        \label{fig:cmap_noise_0.04}
    \end{subfigure}
    \begin{subfigure}[t]{0.15 \textwidth}
        \includegraphics[width=\textwidth]{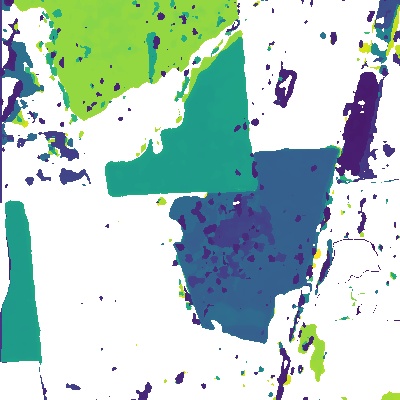}
        \caption*{6\%}
        \label{fig:cmap_noise_0.06}
    \end{subfigure}
    \begin{subfigure}[t]{0.15 \textwidth}
        \includegraphics[width=\textwidth]{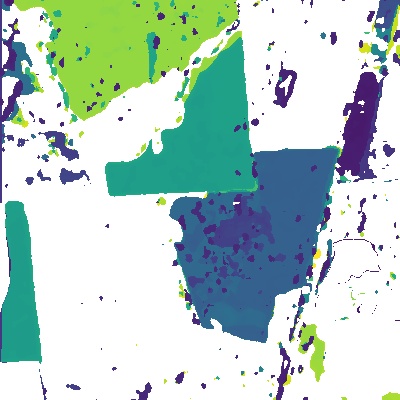}
        \caption*{8\%}
        \label{fig:cmap_noise_0.08}
    \end{subfigure}
    \begin{subfigure}[t]{0.15 \textwidth}
        \includegraphics[width=\textwidth]{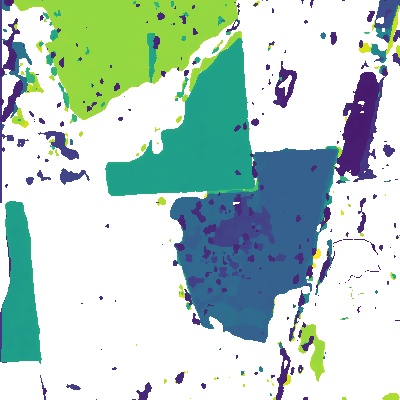}
        \caption*{10\%}
        \label{fig:cmap_noise_0.10}
    \end{subfigure}
    \begin{subfigure}[t]{0.15 \textwidth}
        \includegraphics[width=\textwidth]{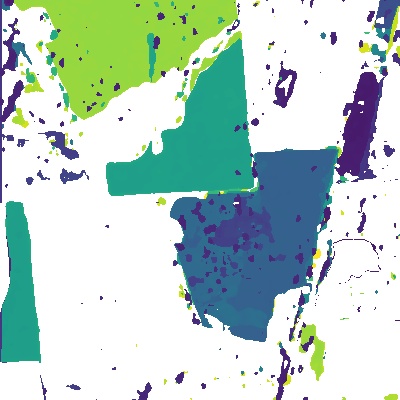}
        \caption*{12\%}
        \label{fig:cmap_noise_0.12}
    \end{subfigure}
    \begin{subfigure}[t]{0.15 \textwidth}
        \includegraphics[width=\textwidth]{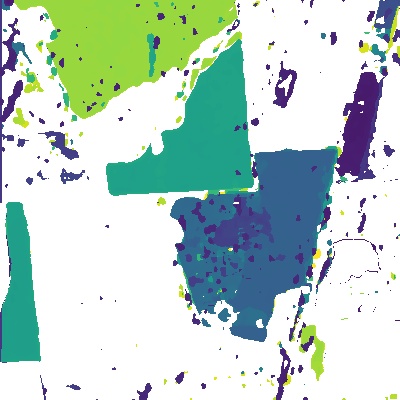}
        \caption*{14\%}
        \label{fig:cmap_noise_0.14}
    \end{subfigure}
    \begin{subfigure}[t]{0.15 \textwidth}
        \includegraphics[width=\textwidth]{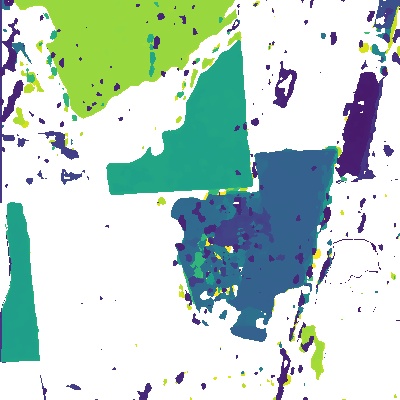}
        \caption*{16\%}
        \label{fig:cmap_noise_0.16}
    \end{subfigure}
    \begin{subfigure}[t]{0.15 \textwidth}
        \includegraphics[width=\textwidth]{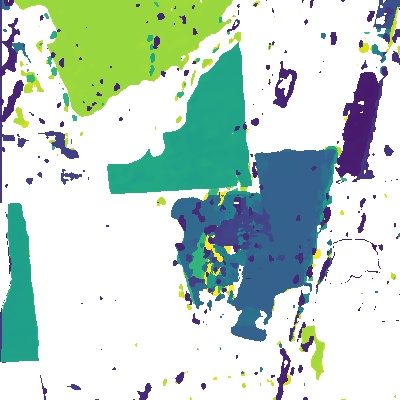}
        \caption*{18\%}
        \label{fig:cmap_noise_0.18}
    \end{subfigure}
    \begin{subfigure}[t]{0.15 \textwidth}
        \includegraphics[width=\textwidth]{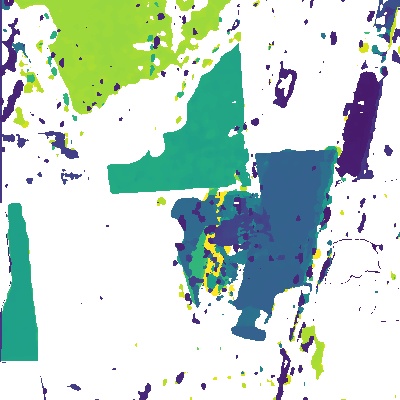}
        \caption*{20\%}
        \label{fig:cmap_noise_0.20}
    \end{subfigure}
    \begin{subfigure}[t]{0.15 \textwidth}
        \includegraphics[width=\textwidth]{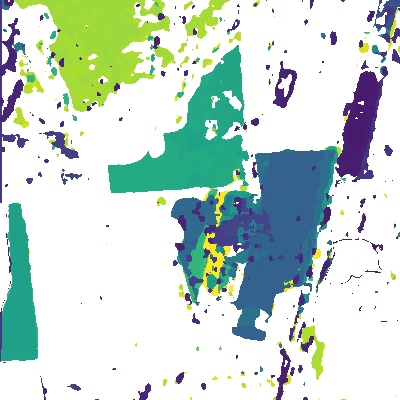}
        \caption*{22\%}
        \label{fig:cmap_noise_0.22}
    \end{subfigure}

    \caption{This figure shows change maps for various levels of training data corruption.
    The percentages indicate the fraction of pixels in the forest reference mask
    that are in reality non-forest pixels.}
    \label{fig:cmaps_noise_fraction}
\end{figure}

\begin{figure}[!ht]
    \centering
    \includegraphics[width=0.7\textwidth]{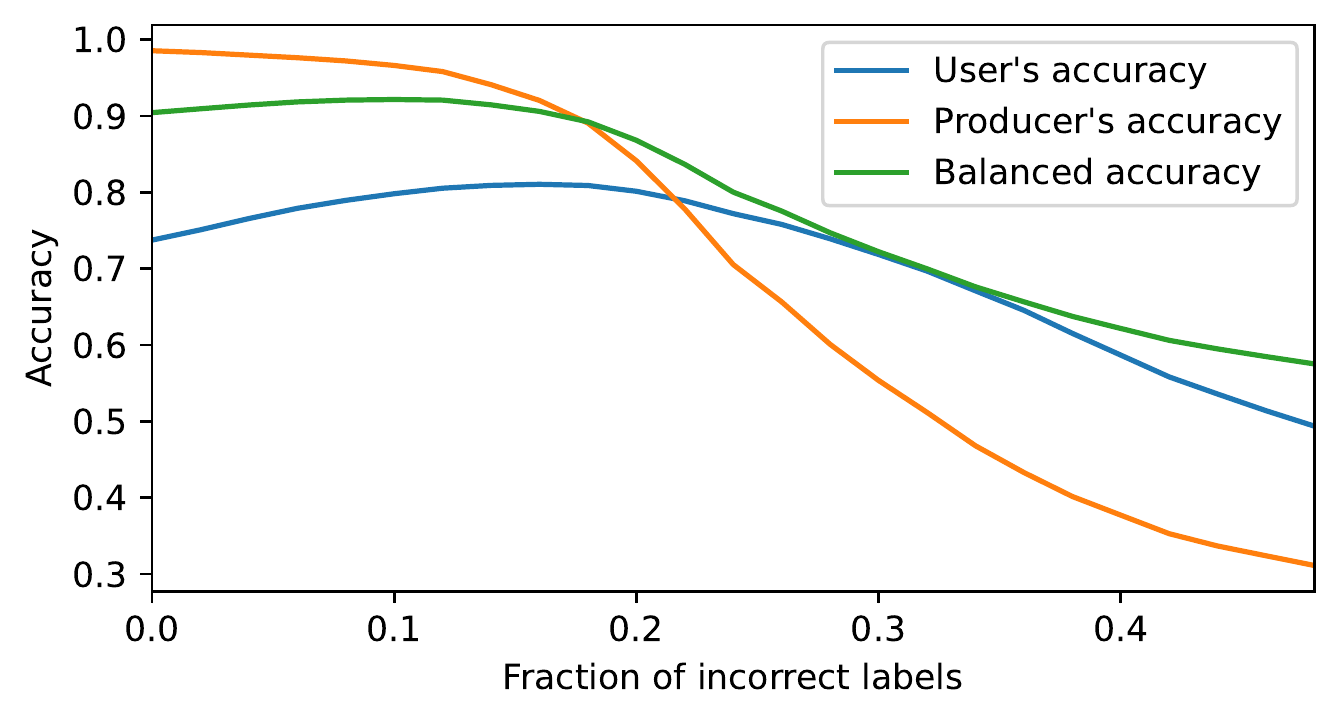}
    \caption{This graph shows the effect of training data corruption on the change detection accuracy.
    The \gls{UA}, \gls{PA}, and \gls{OA} are shown as a function of
    the fraction of incorrect labels.}
    \label{fig:noise_fraction_accuracy}
\end{figure}

\Cref{fig:cmaps_noise_fraction} shows these maps of detected change for fractions between 0\% and 22\%.
Up to about 10\% incorrectly classified pixels,
there is no major visible degradation of the change map.
At higher percentages of corruption,
changes that occur later in the time series are the first to be missed.
However, a large amount of true changes are still detected up to 22\% corruption.
Furthermore, the corruption of the training data only
increases the error of omission
(the number of missed deforestation events),
but not the commission error
(the number of false positives).
These findings are corroborated by a graph of the accuracy scores
(\gls{UA}, \gls{PA}, and \gls{OA})
as a function of the corruption fraction
in \cref{fig:noise_fraction_accuracy}.
The accuracy scores are computed with respect to
the visually interpreted change reference (\cref{fig:reference_visual}).
The \gls{PA} is decreased significantly above 10\% corruption,
but the \gls{UA} does not significantly decrease --
in fact, it increases slightly initially
as the overall number of detected changes is reduced,
including previous false positives.
As a consequence,
the \gls{OA} does not deteriorate significantly even at 20\% corruption.
However,
the \gls{PA} is arguably the most important metric
as the correct identification of all deforestation events
is the primary goal.
As stated previously,
the false positives mostly occur in areas of non-forest and
could be removed by other means,
e.g. by comparing the results with optical data
where non-forest areas may have been apparent even at the start of the time series.

\FloatBarrier
\section{Discussion}


We have shown in the preceding sections
that the proposed method is very successful at picking up deforestation events
in diverse enviroments
by quantifying the deviation from a reference class characterized by a static reference map.
This is true even in adverse conditions such as the site in Indonesia,
where deforestation is hard to make out from Sentinel-1 data even by visual inspection.
Statistically significant change deviations are reliably picked up,
resulting in a small false negative rate.

We have already touched upon possible limitations of the method.
Firstly, there are cases where the backscatter signature of non-forest areas is
very similar to forest areas temporarily.
Nearly all false positives occur in such areas,
when the backscatter is similar to forest at the start of the time series.
Consequently,
all of these false positives are detected at dates near the beginning of the time period.
These commission errors can be reduced by extending the time series into the past
and reserving the first e.g. six months to ascertain which areas were forest initially.

However, much of these shortcoming may in fact be due to the inadequacy of Sentinel-1 data
for the deforestation detection task.
It would therefore be useful to explore this method with additional data sources.
As the proposed method is entirely agnostic to the type of input data,
it can also be applied to other SAR sensors and optical data,
even in combination with Sentinel-1 as it supports multivariate data.

Another issue with this method and Sentinel-1 is the sensitivity to terrain.
As terrain has a big impact on SAR backscatter values,
backscatter deviations from the reference class in areas with uneven topography
may occur due to deviations in terrain rather than deviations in the actual land cover.
This was observed in part of the Indonesian site,
where a mountain in the south western corner caused false positives.
This could potentially be addressed by appropriate terrain flattening,
or including optical data that are less sensitive to terrain.

It is also worth noting that this method is not specific to deforestation,
but can be applied to any use case where changes from some reference class
need to be detected.
This includes applications in agriculture,
e.g. detecting harvesting when only a map of agricultural fields in the region is known.

\section{Conclusions}


The development of this change detection method was motivated by
some of the biggest challenges around deforestation detection using SAR data,
in particular
(1) the lack of reliable reference data and
(2) the fact that changes in backscatter need not represent
changes in land cover/use.
We have shown in this paper that some of these challenges may be addressed
with a semi-supervised approach where changes are detected
with respect to the prototype time series of a reference class.
This prototype time series captures the natural variability of the
forest backscatter signature and removes the potential for
mistakenly detecting changes due to other effects.
In addition,
this method works with noisy reference data as it is robust
to outliers present in the reference class ensemble,
with no significant degradation of the results
up to about 10--20\% reference data corruption.
By detecting changes as deviations from a single reference class rather than
by classifying between two possible classes,
we remove the need to parametrize the non-forest class.
This is advantageous for generalizing the method
as the non-forest class is inherently heterogeneous and may include
many different land cover types.

While the results from this method do not necessarily outperform the reference maps
in terms of spatial accuracy,
they do provide an actual date of change rather than just a year
which is advantageous for near-real-time forest monitoring systems
and critical for actionable insights.
They change maps can also be generated from static reference maps without requiring noise-free reference data.
In this way, the method can be thought of as a means to iteratively improve existing change maps.

In conclusion,
Sentinel-1 SAR data are certainly capable of detecting
deforestation,
even though a longer time series may be required to reliably
tell apart forest and non-forest --
potentially limiting the temporal change detection accuracy.
There is no doubt that more innovations are possible
in the field of multi-sensor fusion --
combining the advantages of SAR, lidar, and optical data --
and the use of spatial features for separating forest and non-forest
with even shorter time series.



\bibliography{references}

\end{document}